\documentclass[aps,pre,twocolumn,superscriptaddress,longbibliography]{revtex4-2}%
\usepackage{amsfonts}
\usepackage{amsmath}
\usepackage{amssymb}
\usepackage{graphicx}
\usepackage{color}
\usepackage{physics}
\usepackage{float}
\usepackage{mathtools}
\usepackage{algorithm}
\usepackage{algpseudocode}
\usepackage{wrapfig}
\usepackage{mathbbol}
\usepackage{upgreek}
\usepackage[normalem]{ulem}

\usepackage[colorinlistoftodos]{todonotes}
\usepackage[colorlinks=true, allcolors=blue]{hyperref}
\usepackage{mwe}
\usepackage{epstopdf}
\usepackage[doipre={DOI:~}]{uri}

\newcommand \be {\begin{eqnarray}}
\newcommand \ee {\end{eqnarray}}
\newcommand \ben {\begin{eqnarray}}
\newcommand \een {\end{eqnarray}}

\newcommand \kv {\mathbf{G}}

\newcommand \dr {\, {\rm d} \mathbf{r}}

\newcommand{\parall}{{||}}
\newcommand{\Gvn}{\mathbf{G}_n}
\newcommand{\Gv}{\mathbf{G}}

\newcommand{\rv}{\mathbf{r}}
\newcommand{\vv}{\mathbf{v}}
\newcommand{\bv}{\mathbf{b}}
\newcommand{\Gvpar}{\mathbf{G}^{\parall}_n}
\newcommand{\Gvperp}{\mathbf{G}^{\perp}_n}
\newcommand{\Gpar}{\mathbf{G}^{\parall}_n}

\newcommand{\Gperpcn}[1]{G_{n,#1}^{\perp}}
\newcommand{\Gparcn}[1]{G_{n,#1}^{\parall}}

\newcommand{\I}{\mathbb{i}}

\begin{document}
\title{Mesoscale Field Theory for Quasicrystals}

\author{Marcello De Donno}
\affiliation{Institute of Scientific Computing, TU Dresden, 01062 Dresden, Germany}
\author{Luiza Angheluta}
\affiliation{Njord Centre, Department of Physics, University of Oslo, 0371 Oslo, Norway}
\author{Ken R. Elder}
\affiliation{Department of Physics, Oakland University, Rochester, Michigan 48309, USA}
\author{Marco Salvalaglio}
\email{marco.salvalaglio@tu-dresden.de}
\affiliation{Institute  of Scientific Computing,  TU  Dresden,  01062  Dresden,  Germany}
\affiliation{Dresden Center for Computational Materials Science (DCMS), TU Dresden, 01062 Dresden, Germany}

\begin{abstract}
We present a mesoscale field theory unifying the modeling of growth, elasticity, and dislocations in quasicrystals. The theory is based on the amplitudes entering their density-wave representation.
We introduce a free energy functional for complex amplitudes and assume non-conserved dissipative dynamics to describe their evolution. 
Elasticity, including phononic and phasonic deformations, along with defect nucleation and motion, emerges self-consistently by prescribing only the symmetry of quasicrystals. Predictions on the formation of semi-coherent interfaces and dislocation kinematics are given.
\end{abstract}
\maketitle

\section{Introduction}

Quasicrystals (QCs) are aperiodic yet ordered arrangements that lack translational symmetry but still possess rotational symmetry. They exhibit distinctive, discrete diffraction patterns, which have been instrumental to their discovery \cite{Levine1984quasicrystals,Shechtman1984metallic} 
and detection in both synthetic and natural materials \cite{Ishimasa1985new,DiVincenzo1991quasicrystals,janot1994quasicrystals,Bindi2009natural}. 
QCs exhibit exotic features such as low friction and thermal conductivity, nonstick surface properties, and peculiar electronic properties \cite{Macia2006,dubois2012properties}.
Importantly, quasicrystalline order can be found in various systems, spanning from solid-state materials to soft matter 
\cite{xianbing2004,Hayashida2007,talapin2009qc,C2CS35063G,Tomonari1197QC,Jules2008archimedean,forster2013quasicrystalline}. Moreover, QCs are intimately related to mathematical tiling concepts explored well before their discovery in actual materials \cite{penrose1974role,Lu2007medieval} and emerge in more exotic systems such as vibrating (macroscopic) granular materials \cite{plati2023quasicrystalline} and quantum phase transitions \cite{Mivehvar2019}. 

QCs can be constructed via different procedures from a periodic hyperlattice \cite{Socoloar1986phonons,Bak86}. Via the \textit{strip-projection} method, for instance, one considers the hyperlattice points within two parallel (flat) hypersurfaces (a hyperstrip) oriented with an irrational slope with respect to the hyperlattice orientations.
The aperiodic structure is obtained by projecting the lattice positions on one of these hypersurfaces.
A classic example is the 1D aperiodic arrangement corresponding to the Fibonacci sequence constructed from a 2D square lattice with this method \cite{guyot2001dislocations}. Similarly, a 2D ten-fold QC \cite{Bendersky1985quasicystal} or the iconic 3D icosahedral QC \cite{Levine1984quasicrystals} can be constructed from periodic lattices in 4D and 6D hyperspaces, respectively. 

A natural description leveraging \textit{discrete diffraction diagrams}, which is thus valid for both periodic crystals and QCs
\footnote{Discrete diffraction diagrams became the defining property of
crystals by the International Union of Crystallography since 1992, fostered by
the discovery of QCs \cite{:es0177}}, 
is obtained via a smooth, dimensionless 
density field $\psi \equiv \psi(\rv)$ 
expanded in \textit{density-waves}
\cite{Levine85elasticity,Socoloar1986phonons}
\begin{equation}\label{eq:psir}
    \psi=\psi_0
    +
    \sum_{n=1}^N \eta_n e^{\I \Gvn \cdot \rv} + \text{c.c.},
\end{equation}
with $\I$ the imaginary unit, c.c. the complex conjugate and $\{\Gvn\}$ the \textit{discrete} set of reciprocal-space 
vectors, at which diffraction peaks are expected \footnote{Note that the formulation in \eqref{eq:psir} with c.c. accounts for $\pm \Gvn$ $\forall n$}. 
The complex amplitude functions,
$\eta_{n}=\phi_{n}e^{\I\theta_n}$,  
are slowly varying (hydrodynamic) fields, and $\psi_0$ is the average density, which here is set to zero for simplicity.
The amplitudes encode lattice deformations through their phases $\theta_n$, defined differently for periodic crystals and QCs, as discussed in the following. Amplitudes remain slowly varying for small deformations, i.e., in elastic regimes and in the presence of isolated defects \cite{salvalaglio2022coarse}.

The density-wave representation \eqref{eq:psir} links directly to Landau’s theories of phase transitions through free energy functionals $F[\psi]$. Free energies for bulk QCs have been discussed in seminal works~\cite{Mermin1985,Lifshitz1997}. 
Approaches like the Swift-Hohenberg model \cite{Swift1977,Cross1993}, the phase field crystal model (PFC) \cite{Elder2002}, as well as the classical density functional theory \cite{Ramakrishnan1979,Vrugt2020}, are based on free energies for smooth density fields where deformations and interfaces can also be described. 
Though primarily applied to ordered and periodic systems, they have yielded remarkable results for QCs too \cite{Barkan2011,Rottler_2012,Archer2013,Achim2014,Barkan2014,xue2022atomic}. 
These methods, however, focus on microscopic length scales, preventing the description of large-scale systems and mechanical properties approaching continuum limits. 

In this work, we introduce and demonstrate a self-consistent mesoscale field theory for QCs. 
This description focuses on complex amplitudes $\{\eta_n\}$, building on coarse-graining concepts introduced for microscopic densities in crystalline systems \cite{Goldenfeld2005,SalvalaglioNPJ2019,salvalaglio2022coarse}. 
Accordingly, it allows one to describe macroscopic aspects, such as different phases and continuum elasticity, while naturally retaining microscopic details, such as the symmetry of QCs and its impact on dislocation nucleation, topological charge, and mechanical properties. 
Although we discuss minimal settings regarding simple QC symmetries and dynamical regimes, the formulation we introduce here can be extended and applied to more complex quasicrystalline systems and refined elasticity models. Still, we explicitly show that the proposed theory already provides novel predictions concerning defect arrangements in QCs. 
We present a free energy for the amplitudes representing the quasicrystalline order in Sect.~\ref{sec:free-energy}. In Sect.~\ref{sec:dislo}, we discuss how such free energy naturally supports the description of dislocations in QCs. Details concerning the underlying elasticity theory are given in Sect.~\ref{sec:elasticity}. We then apply the resulting theory, first to the characterization of dislocation arrangements at a small-angle grain boundary in Sect.~\ref{sec:gbs}, clarifying a scenario that significantly deviates from periodic crystals, and then to the description of defect kinematics in Sect.~\ref{sec:kinematics}. The main conclusions are summarized in Sect.~\ref{sec:conclusions}.

\begin{figure*}
    \centering
    \includegraphics{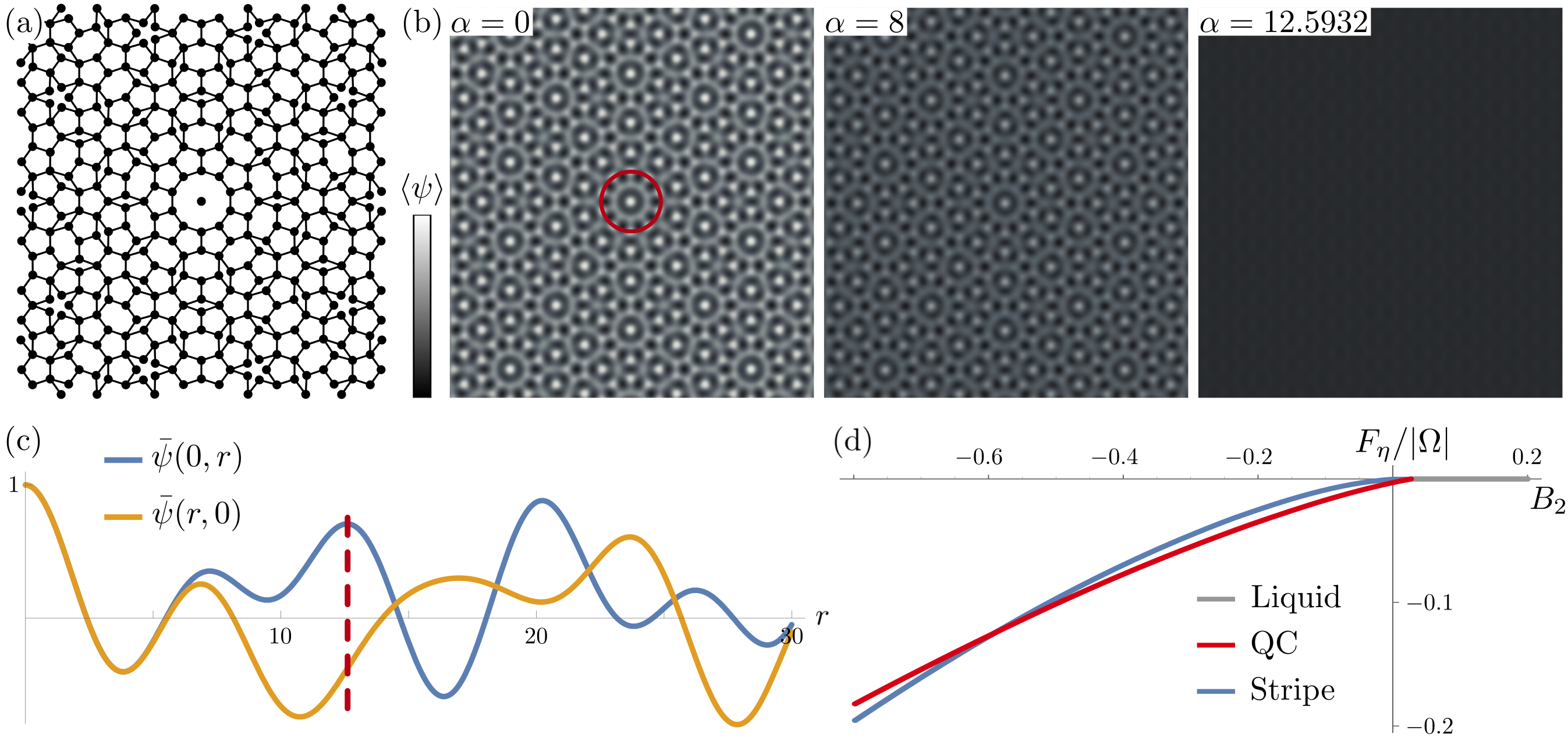}
    \caption{
    \textit{Decagonal QC phase as minimizer of the free energy \eqref{eq:free-energy}}.
    (a) Tiling reconstructed from the density $\psi$ as from Eq.~\eqref{eq:psir} with $\Gvn=\Gvn^{\mathcal{D}}$ and constant, equal (real) amplitudes (shown explicitly in panel (b) for $\alpha=0$). See also the Appendix \ref{app:tiling} for further details on how the tiling is constructed.
    (b) Plots of the coarse-grained density $\langle \psi \rangle$ for different values of the coarsening length $\alpha$ in Eq.~\eqref{eq:cg}. The red circle has a radius of $\Lambda=12.5932$.
    (c) Plots of the normalized density $\bar\psi = \psi / \psi_{\rm max}$ along the directions $x=0$ (blue) and $y=0$ (yellow). The density peak at $y\sim \Lambda$ is highlighted by the red dashed line.
    (d) Free energy density for different phases varying $B_2$ with $B_4=0$, $B_5=-100$, $B_6=0.1$ (independent of $A$). The decagonal QC phase results the minimizer of the free energy \eqref{eq:free-energy} for $-0.592 \lesssim B_2 \lesssim 0.026$.}
    \label{fig:figure1}
\end{figure*}

\section{Free energy}
\label{sec:free-energy}

We start with the Swift-Hohenberg energy functional used in the classical PFC model 
\cite{Elder2002},
\begin{equation}\label{eq:supp:fpfc} 
    F_\psi=\int_{\Omega} \left[\frac{A}{2} \psi(q^2+\nabla^2)^2\psi + \frac{B}{2}\psi^2
        +\frac{C}{3}\psi^3+\frac{D}{4}\psi^4 \right]d\mathbf{r},
\end{equation}
with $q$ controlling the characteristic wave number for minimizers of $F_\psi$ (e.g., enforcing a periodicity of $2\pi/q$ for stripe phases in 2D). 
For periodic crystals, a free energy functional $F_\eta$ depending on amplitudes $\eta_j$ can be derived using a renormalization group approach or simply by expressing $\psi$ via the amplitude expansion \eqref{eq:psir} and integrating over the unit cell \cite{Goldenfeld2005,salvalaglio2022coarse}. It reads as
\begin{equation}\label{eq:supp:feta}
\begin{split}
    F_{\eta,P}
    &=\int_{\Omega} \bigg[ \sum_{n=1}^N A|\mathcal{G}_{n}\eta_n|^2 
    +\sum_{p=2}^P B_p\zeta_p  \bigg]\ d\mathbf{r},
\end{split}
\end{equation}
with $P=4$, $\mathcal{G}_{n}=\nabla^2+2{\rm \I}\kv_n\cdot\nabla$, and the polynomial term of the energy density given by 
\begin{equation}\label{eq:pol_resonance}
\begin{split}
    \zeta_2=&\sum_{p,q}\eta_p\eta_q \delta_{\mathbf{0},\kv_p+\kv_q}
    =2\sum_{n=1}^N |\eta_n|^2=\Phi,
    \\
    \zeta_3=&\sum_{p,q,r}\eta_p\eta_q\eta_r \delta_{\mathbf{0},\kv_p+\kv_q+{\kv_r}},
    \\
    \zeta_4=&\sum_{p,q,r,s}\eta_p\eta_q\eta_r\eta_s \delta_{\mathbf{0},\kv_p+\kv_q+{\kv_r}+\kv_s},
\end{split}
\end{equation}
where the summations are from $-N$ to $+N$ excluding zero, 
$\eta_{-n}=\eta_{n}^*$, and $\Gv_{-n}=-\Gv_{n}$. Explicit expressions for $\zeta_{l}$ for various crystal symmetries can be found in Ref.~\cite{salvalaglio2022coarse}. We remark the presence of the Kronecker delta in these terms, meaning that products of the $l$ amplitudes and their complex conjugate are included if and only if the vector sum of the corresponding reciprocal-space vectors ($\Gv_n$) is zero. 
This is known as resonance condition, where the corresponding product of Fourier modes loses the microscopic periodicity and, therefore, does not cancel out when integrated over the unit cell.

QCs are characterized by aperiodic lattices in direct and reciprocal space. 
For instance, 2D decagonal ($\mathcal{D}$, shown in Fig.~\ref{fig:figure1}a) and 3D icosahedral ($\mathcal{I}$) QCs are well described by the following sets of reciprocal-space vectors \cite{Bendersky1985quasicystal,Levine1984quasicrystals}:
\begin{equation}\label{eq:k-vectors}
\begin{split}
    \Gv_n^{\rm \mathcal{D}}&=[\cos(2\pi n/5),\sin(2\pi n/5)], \\
    \Gv_n^{\rm \mathcal{I}}&=I_0[\cos(2\pi n/5),\sin(2\pi n/5),1/2], \\
    \Gv^{\rm \mathcal{I}}_6&=[0,0,1], 
\end{split}   
\end{equation}
for $1\leq n \leq 5$ with $I_0 = 2/\sqrt{5}$. 
Evidently, these sets of vectors do not form periodic reciprocal lattices, as they feature more independent elements than the dimensionality of the system.
These distinctive characteristics introduce two additional aspects concerning $F_\eta$: 
i) its original derivation relies on the existence of a well-defined unit cell, which does not apply to QCs; 
ii) there is no set of three or four vectors $\Gv_n$ as defined in Eq.~\eqref{eq:k-vectors} for which the resonance conditions in Eq.~\eqref{eq:pol_resonance} are realized, meaning that the QC is not a stable phase according to the energy defined by Eqs.~\eqref{eq:supp:feta} and \eqref{eq:pol_resonance}.

While a unit cell cannot be defined for QCs, quasi-unit-cell descriptions \cite{Socolar86unitcell} featuring overlapping local motifs were proposed and verified experimentally \cite{steinhardt1998experimental}. 
Accordingly, we found that averaging the microscopic density $\psi$ above a characteristic average width $\Lambda$ results in a uniform field, justifying the coarse-graining underlying Eq.~\eqref{eq:supp:feta} for QCs too. 
In particular, a coarse-grained density field $\langle \psi \rangle$ can be evaluated by the Gaussian convolution in 2D \cite{hirvonen_PhysRevB.100.165412,skogvoll2021_10.1103},
\begin{equation}\label{eq:cg}
    \langle \psi \rangle (\mathbf{r}) = 
    \int d\mathbf{r'} \frac{\psi(\mathbf{r'})}{2\pi\alpha^2} 
    \exp\bigg( -\frac{(\mathbf{r}-\mathbf{r'})^2}{2\alpha^2} \bigg), 
\end{equation}
where $\alpha$ is the coarse-graining length. 
Fig.~\ref{fig:figure1}b shows the result of applying Eq.~\eqref{eq:cg} to the density of a decagonal QC for different values of $\alpha$. For $\alpha\sim \Lambda = 12.5932$, relative changes in the macroscopic density decrease by three orders of magnitudes. It thus represents a suitable coarsening length for decagonal QCs. Interestingly, this value corresponds to the distance from the ``center" of the QC to ten symmetric density maxima, as shown in Fig.~\ref{fig:figure1}c (see red dashed line).
The existence of such a (finite) coarsening length justifies the construction of free energy for the amplitudes in analogy to previous works on periodic crystals \cite{salvalaglio2022coarse}. Moreover, this free energy is further shown below to be consistent with several aspects of QCs that are accessible from other theories. We remark that lengths below $\Lambda$ value cannot be described well by coarse-grained approaches, like our framework. On the other hand, no upper bound exists for the validity of the proposed approach. 

The aforementioned issue concerning resonance conditions can be overcome by increasing the degree of the free energy polynomial ($P$). This concept traces back to the first theories of QCs for bulk systems \cite{Mermin1985,Lifshitz1997}. 
For the decagonal or icosahedral quasicrystal, it is enough to consider $P=6$, which corresponds to retaining the next two higher-order terms in the polynomial entering the Swift-Hohenberg energy functional. We remark that the highest order must be even to ensure the existence of a global minimum. 
Explicitly, the free energy then results
\begin{equation}\label{eq:free-energy}
\begin{split}    
    F_{\eta,6} &= \int_{\Omega} \bigg[\sum_{n=1}^N A|\mathcal{G}_{n}\eta_n|^2 + 
    \\&\qquad
    B_2\zeta_2+B_3\zeta_3+B_4\zeta_4+B_5\zeta_5+B_6\zeta_6\bigg]\ d\mathbf{r},
\end{split}
\end{equation}
with $\zeta_5$ and $\zeta_6$ obtained by extending the sums in Eq.~\eqref{eq:pol_resonance} to products of five and six amplitudes, respectively. For decagonal QCs, the newly introduced terms read: 
\begin{equation}\label{eq:supp:zeta_5_6}
    \begin{split}
            \zeta_5^{\mathcal{D}} &= \bigg(\prod_{j=1}^5 \eta_j\bigg)+\mathrm{c.c.}, \\
            \zeta_6^{\mathcal{D}} &= 720 \sum_{\substack{i\\j>i\\k>j}} |\eta_i|^2|\eta_j|^2|\eta_k|^2 + \\
            & \qquad 180\sum_{\substack{i\\j\neq i}}|\eta_i|^4|\eta_j|^2
                 +20\sum_i|\eta_i|^6. 
    \end{split}
\end{equation}
For icosahedral QCs, there are no combinations of three or five $\Gv_n^{\mathcal{I}}$ with zero sum, meaning that $\zeta_3^{\mathcal{I}} = \zeta_5^{\mathcal{I}} = 0$. The explicit definition of $\zeta_6^{I}$ follows by considering the expression for $\zeta_6^{D}$ in Eq.~\eqref{eq:supp:zeta_5_6} with an additional term reading $(\prod_{j=1}^6 \eta_j)+\mathrm{c.c.}$, taking into account that the sum of the six vectors $\Gv_n^{\mathcal{I}}$ is zero.
Although not specifically addressed here, we expect that for higher polynomial degrees $P$, the proposed free energy may describe even more complex QC symmetries as stable phases.

Fig.~\ref{fig:figure1}d shows that a decagonal QC phase minimizes the free energy (\ref{eq:free-energy}) for some parameters. We vary $B_2$, corresponding to a phenomenological temperature parameter in analogy with classical PFC models \cite{Elder2002}. Consistently, for large $B_2$, disordered/liquid phases are favored, while first QCs and then stripe phases minimize the free energy when decreasing $B_2$. 
By considering $\psi_0 \neq 0$ and spatially dependent, the theory can be straightforwardly extended to admit phase coexistence \cite{Yeon2010,salvalaglio2022coarse}, whose discussion is however beyond the scope of the present work.

In its simplest form, the dynamics of the order parameter in the PFC model is given by the conservative evolution law 
\begin{equation}\label{eq:dpsidt}
    \frac{\partial \psi}{\partial t}=\nabla^2 \frac{\delta F_\psi}{\delta \psi}.
\end{equation}
Under similar assumptions underlying the derivation of $F_{\eta,P}$ \cite{salvalaglio2022coarse},
the evolution law for amplitudes approximating Eq.~\eqref{eq:dpsidt} is
\begin{equation}\label{eq:dynamicseta}
\frac{\partial \eta_j}{\partial t}=-|\kv_j|^2\frac{\delta F_{\eta,P}}{\delta \eta_j^*}=-|\kv_j|^2 (
A \mathcal{G}_n^2
    \eta_n + \sum_{p=2}^P B_p \partial_{\eta_n^*} \zeta_p),
\end{equation}
which describes the evolution of a QC similarly to periodic crystals. An additional timescale to properly account for elastic relaxation may also be considered, leveraging the hydrodynamic formulation introduced in Ref.~\cite{Heinonen2016} that is compatible with the proposed free energy. We expect this extension to be relevant for fast dynamics and theoretical analysis of competitive relaxation mechanisms. However, focusing here on assessing the fundamental aspects of the proposed theory, we refrain from considering such an extension while targeting it in future works. 
Numerical examples are reported in the next sections, solving Eq.~\eqref{eq:dynamicseta} via a standard Fourier pseudo-spectral method for spatial discretization \cite{salvalaglio2022coarse}. Further details can be found in Appendix \ref{app:numerics}. The datasets generated and analyzed during the current study are openly available in Ref.~\cite{DataSet}.

\section{Growth and Dislocations}
\label{sec:dislo}

The free energy, Eq.~\eqref{eq:free-energy}, and the corresponding dynamics, Eq.~\eqref{eq:dynamicseta}, enable the study of out-of-equilibrium settings, their evolution, and the deformation of QCs. 
Moreover, deformations and defects in QCs can be fully characterized via the phase of the complex amplitudes $\{\eta_n\}$.

We recall that, unlike periodic crystals, the full description of the order realized in QCs requires the definition of a periodic lattice in a higher-dimensional space \cite{Socoloar1986phonons,Bak86}. Namely, a QC in 2D is represented by a periodic hyper-lattice $\mathcal{L}$ in a 4D hyper-space with coordinates $\widetilde{\rv}=\rv^{\parall} \oplus \rv^{\perp}$, with $\rv^{\parall}=\rv$ the coordinates the so-called \textit{parallel} space ($\Omega^{||}=\Omega$), the physical space of definition of the quasicrystal, and $\rv^{\perp}$ the coordinates of the so-called \textit{perpendicular} space ($\Omega^{\perp}$)\cite{guyot2001dislocations}, required to define $\mathcal{L}$ in addition to $\Omega^{||}$. 

Elastic deformation of $\mathcal{L}$ can be generally described by the displacement field 
$\widetilde{\mathbf{U}}=\mathbf{u} \oplus \mathbf{w}$ with $\mathbf{u}$ corresponding to the displacements in $\Omega^{||}$, also called \textit{phonons}, and $\mathbf{w}$ being the displacements in $\Omega^{\perp}$, 
called \textit{phasons}.
From the deformation of the periodic density $\widetilde{\psi}(\widetilde{\rv}-\widetilde{\mathbf{U}})$ of $\mathcal{L}$, the QC density $\psi=\widetilde{\psi}(\widetilde{\rv})|_{\rv^{\perp}=\mathbf{0}}$ results 
\begin{equation}\label{eq:phases} 
\begin{split}
    \psi&=\sum_{n=1}^N \underbrace{\phi_n e^{ - {\I}\theta_n}}_{\eta_n} e^{{\I}{
    \Gv}_n \cdot {\rv}}+\text{c.c.}, \\
    \theta_n&=\arg(\eta_n)=\Gv_n^{\parall} \cdot \mathbf{u}+\Gv_n^{\perp} \cdot
    \mathbf{w}, 
\end{split} 
\end{equation} 
with $\Gv_n^{\parall}=\Gv_n$ and $\Gvperp$ can be constructed from $\Gvpar$ according to the QC rotational symmetry. 
Following Refs.~\cite{Levine85elasticity,Socoloar1986phonons} for the decagonal QC we set $\Gv_n^{\perp}=a\Gv_{(3n\, {\rm mod}\, 5)}^{\parall}$ with $a=(1+\sqrt{5})/2$.  
A refresher of the construction leading to the definition of direct- and reciprocal-space vectors for decagonal quasicrystals is reported in Appendix~\ref{app:geo} for completeness. 

Note that the phase $\theta_n$ depends on both deformations $\mathbf{u}$ and $\mathbf{w}$. 
Dislocations in QCs inherently induce both phononic and phasonic deformations \cite{Levine85elasticity,Socoloar1986phonons} as they correspond to topological defects in the phase $\theta_n$ with the topological charge given by 
\begin{equation}\label{eq:ointphase} 
\oint d\theta_n = - 2\pi s_n = -(\Gvpar
\cdot \bv^{\parall}+\Gvperp \cdot \bv^{\perp}), 
\end{equation}
$s_n$ the winding (integer) number, $\bv^{\parall}=\oint {\rm d}\mathbf{u}$ and $\bv^{\perp}=\oint {\rm d}\mathbf{w}$ Burgers vectors in $\Omega^{||}$ and $\Omega^{\perp}$, respectively. 

\begin{figure}
    \centering
    \includegraphics{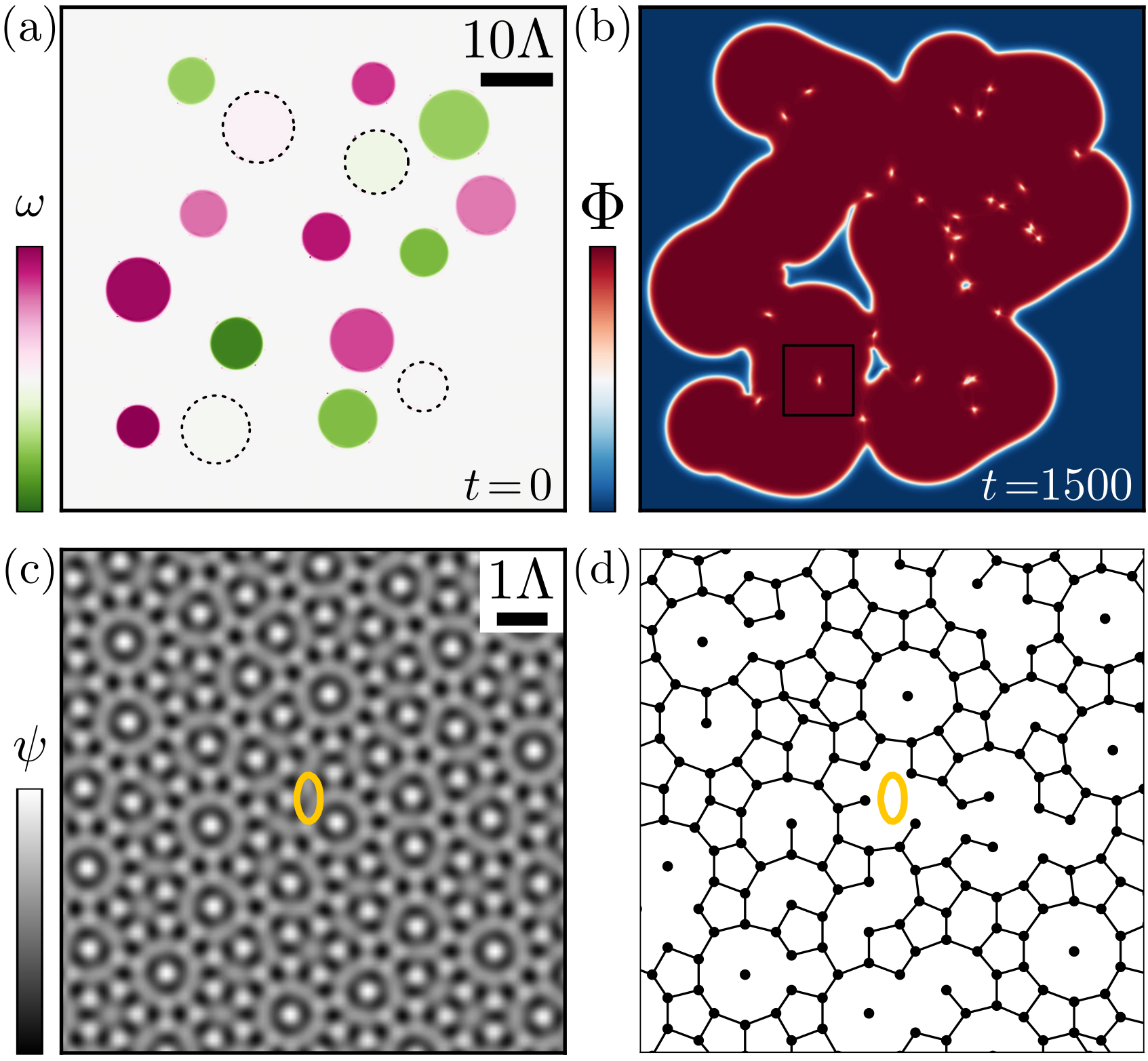}
    \caption{
    \textit{Growth of QCs seeds and defect nucleation}. 
    (a) Rotation field $\omega$ at $t=0$ initialized via amplitudes $\eta_n=\phi_n\text{exp}(\I  (M(\beta)\Gpar-\Gpar) \cdot \rv)$ with $M(\beta)$ the standard rotation matrix \cite{salvalaglio2022coarse} and $|\beta| \leq 5^\circ$. Grains with a small rotation are illustrated by dashed lines.
    (b) Representative stage of growth, illustrated by $\Phi$. 
    (c) and (d): $\psi$ and tiling in the region marked by the black square in panel (b) 
    with a (yellow) isoline at $\Phi/\Phi_{\rm max} = 0.7$ showing the defect location. Parameters as in Fig.~\ref{fig:figure1}d with $B_2=0.02$ and $A=1$.
    }
    \label{fig:figure2}
\end{figure}

An out-of-equilibrium system is illustrated in Fig.~\ref{fig:figure2} by a numerical simulation. 
Slightly misoriented  quasicrystalline seeds  (i.e., rotated by some small angle $\beta$) 
are considered; see Fig.~\ref{fig:figure2}a. Using parameters that favor a QC phase, the initial seeds grow and eventually merge with the formation of topological defects (Fig.~\ref{fig:figure2}b), as indicated by localized regions where $\Phi$ decreases, pointing to a loss of quasicrystalline order. Figures \ref{fig:figure2}c and \ref{fig:figure2}d show the reconstructed density $\psi$ and the tiling around a defect, deviating from the bulk, unperturbed phase (see Fig.~\ref{fig:figure1}). 
The proposed framework thus allows for the description of not only bulk systems but also complex scenarios in out-of-equilibrium settings, including interfaces, deformations, and defects. 
Note that $\Phi$ also enables easy detection of defects in QCs, not immediately accessible from the density or tiling considered by other theories.

\section{Elasticity}
\label{sec:elasticity}

\begin{figure*}
    \centering
    \includegraphics{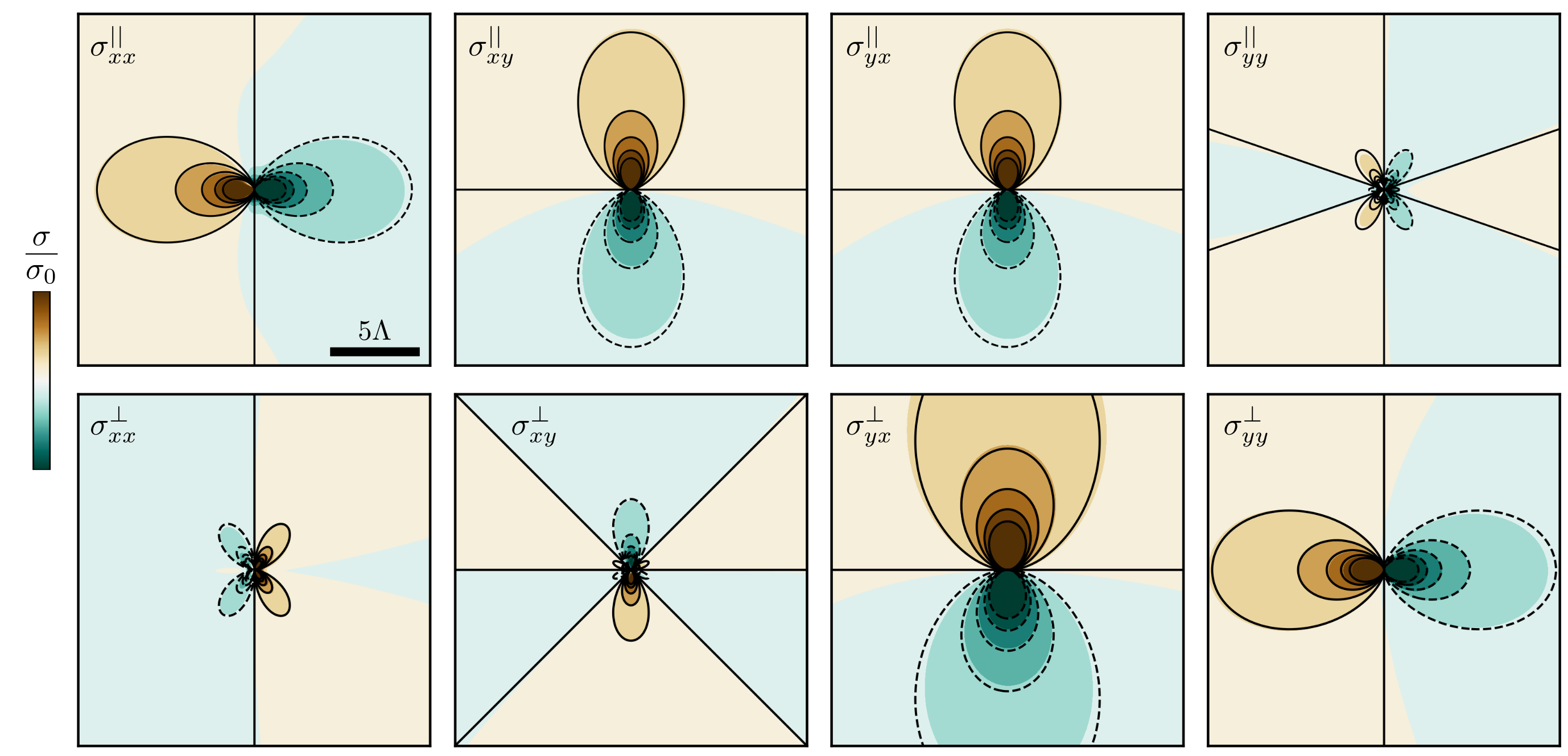}
    \caption{
    \textit{Stress field of a dislocation in a decagonal QC.} 
    Numerical stress field components (Eq.~\eqref{eq:stressAPFC}, filled contour plot) are compared with the analytic stress fields (reported in Appendix \ref{app:analyticstress}, black isolines) for a dislocation with Burgers vectors 
    $\mathbf{b}^\parall = \mathbf{b}^\perp = \left[0,-\frac{8}{5}\pi\sin(\frac{4}{5}\pi)\right]$.
    Dashed isolines correspond to negative values.
    Simulation parameters are set as in Fig.~\ref{fig:figure2}.
    }
    \label{fig:stressComp}
\end{figure*}

A self-consistent elasticity theory for QCs follows upon the deformation of $\psi$. The elastic energy, including all terms depending on deformation gradients, is
\begin{equation}\label{eq:elastic_energy}
\mathcal{E} = \int e(\nabla\mathbf{u},\nabla \mathbf{w}) \dr 
= \int  A \sum_{n=1}^N |\mathcal G_n \eta_n|^2 \dr, 
\end{equation}
as can be obtained by considering the polar representation of complex amplitudes, with phase $\theta_n$ defined in Eq.~\eqref{eq:phases}, in the free energy $F_{\eta,P}$.
The stress fields can be computed from $\{\eta_n\}$ by taking the variational of $\mathcal{E}$ with respect to independent variations of the displacements, 
$\delta \mathcal{E} = \int (\sigma_{ij}^{\parall}\partial_j\delta u_i+\sigma_{ij}^{\perp}\partial_j\delta w_i) \dr$, 
where summation over repeated indices is implied. In terms of amplitudes, $\delta \mathcal{E}$ becomes
\begin{equation}\label{eq:var_elast_phase}
    \delta\mathcal E = 4 A\sum\limits_{n=0}^4 \int d\mathbf r \,
    \Im\left[(\mathcal G^*_n\eta^*_n)(\mathcal Q_{n,j}\eta_n)\right](\partial_j\delta\theta_n).
\end{equation}
See also Appendix \ref{app:energystress} for its step-by-step derivation.
We remark that the expression above is general and, in analogy with the amplitude expansions \eqref{eq:psir}, it holds for both periodic crystals and QCs. For the latter, using Eq.~\eqref{eq:phases} to expand the amplitudes' phase $\theta_n$, we obtain
\begin{equation}
\begin{split}
    \delta\mathcal E &= 4 A\sum\limits_{n=0}^4 \int d\mathbf r 
    \,\Im\left[(\mathcal G^*_n\eta^*_n)(\mathcal Q_{n,j}\eta_n)\right] \times 
    \\&\qquad\qquad\qquad\quad
    \bigg(G^\parall_{n,i}\partial_j\delta u_i+G^\perp_{n,i}\partial_j\delta w_i\bigg),
\end{split}
\end{equation}
where $\delta \mathbf{u}$ and $\delta \mathbf{w}$ are the infinitesimal variations of the displacements in $\Omega^{\parall}$ and $\Omega^{\perp}$ subspace respectively. Thus, following from the definition of stress fields $\sigma_{ij}^{\parall}=\delta F / \delta (\partial_j u_i)$ and $\sigma_{ij}^{\perp}=\delta F / \delta (\partial_j w_i)$, we obtain

\begin{equation}\label{eq:stressAPFC}
\begin{split}
    {\sigma_{ij}^{\rm S}} &= 4 A \sum_{n=1}^N  G^{\rm S}_{n,i}{\rm Im}\left[(\mathcal G^*_n\eta^*_n)((\partial_j +{\I} \Gparcn{j})\eta_n)\right] \\
    & \approx 8 \, \phi^2_0 A \sum_{n=1}^N G^{\rm S}_{n,i}\Gparcn{j}\Gparcn{k}\partial_k\theta_n,
\end{split}
\end{equation}
with $S=(\parall,\perp)$ and $\phi_0$ the (real) value of amplitudes in the bulk. The lowest order term reduces to the linear stress-strain relation. 

For small distortions, the elastic energy density of QCs reduces to the quadratic form \cite{DingPRB93},
\begin{equation}\label{eq:elast_form}
\begin{split}
    2e(\nabla\mathbf{u},\nabla\mathbf{w})=&
    C_{ijkl}\varepsilon_{ij}\varepsilon_{kl} +
    K_{ijkl}\partial_j w_i\partial_{l} w_k +\\
    &R_{ijkl}\varepsilon_{ij}\partial_l w_k +
    R_{ijkl}^\prime\partial_j w_i\varepsilon_{kl},
\end{split}
\end{equation}
with $\varepsilon_{ij}= \frac{1}{2}(\partial_iu_j + \partial_ju_i) \equiv \varepsilon_{ij}^{\parall}$ and $\partial_iw_j\equiv \varepsilon_{ij}^{\perp}$ the phononic and phasonic strains, respectively. 
The resulting constitutive relations are \begin{equation}\label{eq:stressANL}
\begin{split}    
    \sigma_{ij}^\parall &= C_{ijkl} \varepsilon_{kl}^{\parall} +  R_{ijkl} \varepsilon_{kl}^{\perp},\\
    \sigma_{ij}^\perp   &= R_{ijkl}^\prime \varepsilon_{kl}^{\parall} +  K_{ijkl} \varepsilon_{kl}^{\perp}.   
\end{split}
\end{equation}
By comparing these expressions for the stress field or the elastic energy to their counterparts depending on amplitudes, Eqs.~\eqref{eq:stressAPFC} and \eqref{eq:elast_form}, the elastic constants $\mathbf{C}$, $\mathbf{K}$, and
$\mathbf{R}^\prime$=$\mathbf{R}^{\rm T}$ thus result
\begin{equation}\label{eq:elconstants}
\begin{split}
C_{ijkl} &= 8 A\phi_0^2\sum_{n=1}^N  \Gparcn{i}\Gparcn{j}\Gparcn{k}\Gparcn{l},\\
R_{ijkl} &= 16 A\phi_0^2\sum_{n=1}^N  \Gparcn{i}\Gparcn{j}\Gparcn{k}\Gperpcn{l}, \\
K_{ijkl} &= 16 A\phi_0^2\sum_{n=1}^N  \Gparcn{i}\Gperpcn{j}\Gparcn{k}\Gperpcn{l},
\end{split}
\end{equation}
consistent with known results for QCs \cite{Levine85elasticity,DingPRB93}. 
For instance, for decagonal QCs, $C_{ijkl}=\lambda \delta_{ij}\delta_{kl}+
2\mu(\delta_{ik}\delta_{jl}+\delta_{il}\delta_{jk})$ with $\mu$=$\lambda$=$5A\phi^2_0$ (isotropic, with a ratio close to experiments, e.g. for Al–Ni–Co QCs \cite{Chernikov1998}),
$K_{ijkl}=K_1\delta_{ik}\delta_{jl}+K_2(\delta_{ij}\delta_{kl}-\delta_{il}\delta_{jk})$
with $K_1=10(3+\sqrt{5})A\phi^2_0$ and $K_2=0$, and
$R_{1111}=R_{1122}=-R_{2211}=-R_{2222}=R_{1221}=R_{2121}=-R_{1212}=-R_{2112}=R=5(1+\sqrt{5})A\phi^2_0$,
and 0 elsewhere \cite{DingPRB93}. Additionally, inverting the expression of $\theta_n$ in Eq.~\eqref{eq:phases} allows us to determine directly phononic and phasonic strains from a set of complex amplitudes extending the formalism introduced in \cite{SalvalaglioNPJ2019}. The corresponding equations are reported in the Appendix \ref{app:strain}.

Similarly to the classical PFC model, the considered formulation shows limited control of the model parameter over the elastic constants. 
We remark that an extended set of elastic constants enter a more general elastic energy formulation for QCs \cite{QCsmallGB_PhysRevLett.62.2699}.
Furthermore, the proposed theory can be extended towards different elastic materials upon considering more length scales in the differential operator in Eq.~\eqref{eq:free-energy} \cite{Mkhonta2013} or replacing it with the definition of a correlation function (analogously to the so-called structural PFC model \cite{Greenwood2010}). 
Such extensions are indeed compatible with coarse-graining and phase-stability concepts discussed in Sect.~\ref{sec:free-energy}, as well as a description of defects and out-of-equilibrium scenarios presented in Sect.~\ref{sec:dislo}. 

With the quantities derived in this section, we may assess the deformation of QCs described by amplitudes with known results from continuum mechanics. 
For the latter, we can consider the displacement field of a dislocation in an elastically-isotropic QC from Ref.~\cite{PialiPRB1987}, its derivation to obtain the analytic strain field, and then calculate the associated stress field via Eq.~\eqref{eq:stressANL}. 
We refer to such stress field, explicitly reported in Appendix \ref{app:analyticstress}, as \textit{analytic}, in contrast to the \textit{numerical} stress field computed from amplitudes via Eq.~\eqref{eq:stressAPFC}. 
We consider in particular a dislocation dipole, with Burgers vectors $\mathbf{b}^\parall = \mathbf{b}^\perp = \left[0,\pm \frac{8}{5}\pi\sin(\frac{4}{5}\pi)\right]$. This can be simulated by setting displacement field in the initial condition for amplitudes accordingly \cite{salvalaglio2022coarse}. In Fig.~\ref{fig:stressComp}, we show the \textit{numerical} stress field of one of these dislocations. 
Peculiar features observed for the amplitude description of dislocations in crystals, such as an inherent regularization of the elastic field at the dislocation core \cite{SalvalaglioJMPS2020,Benoit-Marechal_2024} are observed here as well. Moreover, the computed fields match almost perfectly the isolines corresponding to the \textit{analytic} stress produced by the same configuration in the far field.

\begin{figure}
    \centering
    \includegraphics{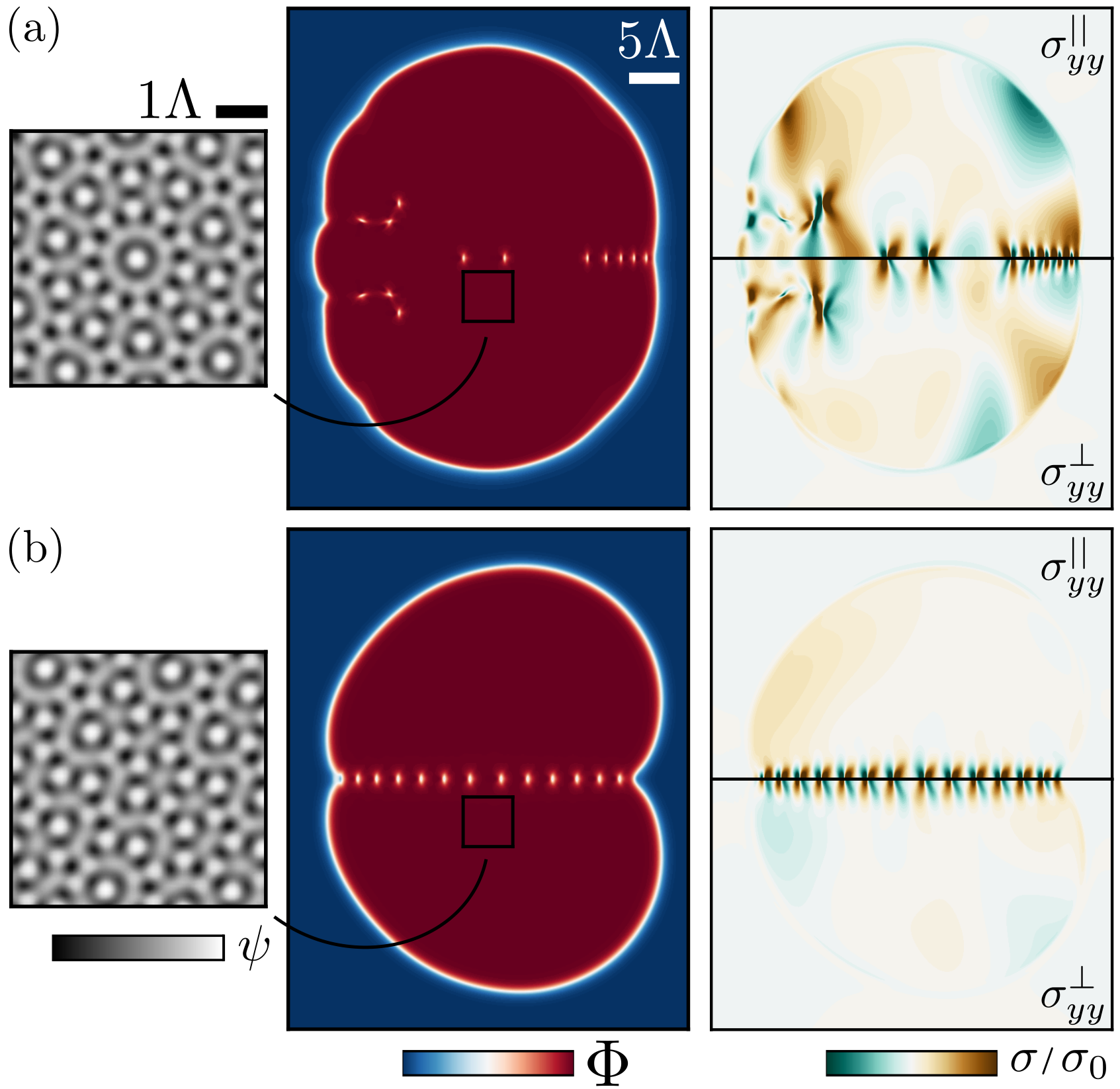}
    \caption{\textit{Growth of QCs and semicoherent interfaces}. 
    (a) Representative stage of the growth of two misoriented circular seeds ($\beta=\pm4^\circ$ set as in Fig.~\ref{fig:figure2}) with radius $4 \Lambda$ and center-to-center distance of $9 \Lambda $ along the y-axis.
    (b) Growth as in (a) with an additional rotation $-3\beta$ in $\Omega^{\perp}$. Parameters as in Fig.~\ref{fig:figure2}. See also Supplemental Videos \cite{SuppVideo}.
    }
    \label{fig:figure3}
\end{figure}

\section{Insights on small-angle grain boundaries}
\label{sec:gbs}

The theory outlined in previous sections allows us to inspect non-trivial settings involving the interplay of growth and dislocation nucleation. In particular, we highlight here the description of defects and deformations during the growth and impingement of two slightly misoriented QCs (rotated $\pm \beta$), recently investigated in experiments \cite{han2021formation}. A representative stage during growth is illustrated in Fig.~\ref{fig:figure3}a. 
From the phases $\theta_n$, we can determine the (4D) Burgers vectors $\widetilde{\mathbf{b}}=\mathbf{b}^{\parall} \oplus \mathbf{b}^\perp$. Defects with (five) different orientations form, having two Burgers vector lengths: 
$|\widetilde{\mathbf{b}}_1|^2= \frac{16}{25}(5-\sqrt5)\pi^2$ for the lowest-energy defects, and 
$|\widetilde{\mathbf{b}}_2|^2= \frac{24}{25}(5-\sqrt5)\pi^2$ for the second-lowest energy.
Moreover, stress fields computed via Eq.~\eqref{eq:stressAPFC} are consistent with the ones obtained by deriving analytic displacements \cite{PialiPRB1987} and multiplying by elastic constants \eqref{eq:elconstants} for the same Burgers vectors (see also Appendix \ref{app:analyticstress}). 

For periodic crystals, a straight semi-coherent interface hosting dislocations of the same kind is expected in the setting of Fig.~\ref{fig:figure3}a (see, e.g., Ref.~\cite{Praetorius2019}). 
A significantly different scenario thus emerges for QCs. Inherent phasonic deformations, absent in the initial rotation, are induced by defects as described by Eq.~\eqref{eq:phases}. Moreover, a rotation by $-3\beta$ in $\Omega^{\perp}$ would be required to rotate the subspaces in a synchronized manner; we recall that $\Gv_n^{\perp} \propto \Gv_{(3n\, {\rm mod}\, 5)}^{\parall}$. 
A $\Omega^{\parall}$ rotation of the QCs thus introduces a geometric frustration, accommodated by the nucleation of defects of different kinds and arranged over a more complex network \cite{QCsmallGB_PhysRevLett.62.2699}. 
For comparison, Fig.~\ref{fig:figure3}b shows that a straight, semi-coherent interface composed of defects of the same kind is indeed obtained by the additional
$-3\beta$ rotation in $\Omega^{\perp}$. However, this setting does not correspond to a physical rotation of the QCs in $\Omega^{\parall}$, as it is visible in the inset in Fig.~\ref{fig:figure3}b, which deviates from a simple rotation of the structure in Fig.~\ref{fig:figure1}b.
Due to such a nontrivial phononic-phasonic deformation, the orientations of the QCs are also varying in space and time, qualitatively reproducing the evidence in \cite{han2021formation}, and nucleation of additional defects at the surface occur at later stages, reminiscent of rearrangements mediated by phasons \cite{Achim2014,NagaoPRL2015}.

\section{Defect Kinematics}
\label{sec:kinematics}

Finally, we show that the proposed field theory also captures self-consistently the driving force for dislocation motion. 
Following the theoretical framework proposed for crystals   \cite{Skaugen2018b,salvalaglioPRL2021,SKOGVOLL2022104932}, we can track Burgers vector densities $\mathbf{B}^{{\rm S}}$ for a dislocation at $\rv_0$ via the zeros of $\eta_n$ corresponding to singularities in the phases $\theta_n$.
In particular, we may express $\mathbf{B}^{{\rm S}}$ as superposition of Dirac-delta distributions $\delta(\eta_n)$, 
\begin{equation}\label{eq:dislodensity}
\begin{split}
    \mathbf{B}^{||}&=\bv^{{\rm ||}}\delta (\rv-\rv_0)= -\frac{4\pi}{N|\bv^{||}|^2}\sum\limits_{n=1}^N \Gv^{{\rm ||}}_n D_n \delta(\eta_n), 
    \\
    \mathbf{B}^{\perp}&=\bv^{{\rm \perp}}\delta (\rv-\rv_0)= -\frac{4\pi}{Na^2|\bv^{\perp}|^2}\sum\limits_{n=1}^N \Gv^{{\rm \perp}}_n D_n \delta(\eta_n),
\end{split}
\end{equation}
following from Eq.~\eqref{eq:ointphase} upon contracting with $\Gv^{{\rm S}}_n$. 
$D_n = \frac{\epsilon_{jk}}{2\I}\partial_j\eta_n^* \partial_k \eta_n $ is the determinant of the coordinate transformation from $\rv$ to [Re($\eta_n$),Im($\eta_n$)]
\footnote{This holds true in 2D, see Ref~\cite{SKOGVOLL2022104932} for extensions to 3D}.
$\mathbf{B}^{\rm S}$ and $D_n$ follow the continuity equations 
\begin{equation}\label{eq:continuity}
\begin{split}
    \partial_t D_n + \partial_j J_{n,j}^{\rm D} &= 0,
    \\
    \partial_t B_i^{\rm S} + \partial_j J_{i,j}^{\rm B,S} &= 0,
\end{split}
\end{equation}
with current densities \cite{SkogvollNPJ2023}
\begin{equation}\label{eq:currentJ}
\begin{split}
    J_{n,j}^{\rm D} &= \epsilon_{jk} \text{Im}(\partial_t\eta_n\partial_k\eta_n^*),
    \\
    J_{i,j}^{\rm B,S} &= b_i^{\rm S} v_j^{\rm S} \delta(\rv-\rv_0),
\end{split}
\end{equation}
where $\mathbf{v}^{\rm S}$ is the dislocation velocity. 
Combining Eqs.~\eqref{eq:dislodensity} and \eqref{eq:currentJ}, we can determine a general expression for the dislocation velocity in each subspace from $\mathbf J_n^{\rm D}$ and thus depending on amplitudes $\{\eta_n\}$ as well as their time evolution ($N=5$ hereafter),
\begin{equation}\label{eq:velocity_diff}
\begin{split}
    v_j^\parall &= \frac{2}{5|\mathbf b^\parall|^2} \times \\ 
    &\sum\limits_{n=1}^5 \bigg( 
    (\mathbf b^\parall\cdot\mathbf G^\parall_n)^2 + 
    (\mathbf b^\parall\cdot\mathbf G^\parall_n) 
    (\mathbf b^\perp\cdot\mathbf G^\perp_n)
    \bigg)\frac{J_{n,j}^{\rm D}}{D_n}\Big|_{\mathbf{r}=\mathbf{0}}, 
    \\
    v_j^\perp &= \frac{2}{5a^2|\mathbf b^\perp|^2} \times \\
    &\sum\limits_{n=1}^5 \bigg( 
    (\mathbf b^\perp\cdot\mathbf G^\perp_n)^2 + 
    (\mathbf b^\parall\cdot\mathbf G^\parall_n) 
    (\mathbf b^\perp\cdot\mathbf G^\perp_n)
    \bigg)\frac{J_{n,j}^{\rm D}}{D_n}\Big|_{\mathbf{r}=\mathbf{0}}.
\end{split}
\end{equation}
As $\bv^{\perp} \cdot \Gvperp = 2\pi - \bv^{\parall} \cdot \Gvpar$ (Eq.~\eqref{eq:ointphase} with $s_n=1$), the defect velocities in each subspace are not independent, but rather constrained by
\begin{equation}\label{eq:vel_constraint}
    \vv^{\parall} |\bv^{\parall}|^2 + \vv^{\perp} a^2|\bv^{\perp}|^2 = 
    \frac{8\pi^2}{5}\sum_{n=1}^5 \frac{\mathbf{J}_{n}^{\rm D}}{D_n}. 
\end{equation}
Furthermore, the geometrical condition 
\begin{equation}\label{eq:geocond}
\begin{split}    
    \frac{1}{|\mathbf b^\parall|^2} \bigg( 
    (\mathbf b^\parall\cdot\mathbf G^\parall_n)^2 + 
    (\mathbf b^\parall\cdot\mathbf G^\parall_n) 
    (\mathbf b^\perp\cdot\mathbf G^\perp_n)\bigg) = 
    \\
    \frac{1}{a^2|\mathbf b^\perp|^2} \bigg( 
    (\mathbf b^\perp\cdot\mathbf G^\perp_n)^2 + 
    (\mathbf b^\parall\cdot\mathbf G^\parall_n) 
    (\mathbf b^\perp\cdot\mathbf G^\perp_n)\bigg),
\end{split}
\end{equation}
$\forall n \in \{1,2,...,5 \}$, is sufficient for these velocities to be equal. We note that having equal velocities in the two subspaces has often been assumed in the modeling of dislocation motion in QCs \cite{Lubensky86dislomotion}. Here, specific conditions are identified to ensure or predict that this identity holds. 
In our numerical experiments (e.g., Figs. \ref{fig:figure2} and \ref{fig:figure3}), we observed the nucleation of dislocations with the lowest and second-lowest energy, for which  
$|\mathbf{\mathbf b^\parall} \oplus \mathbf{\mathbf b^\perp}|^2= \frac{16}{25}(5-\sqrt5)\pi^2$ and  
$|\mathbf{\mathbf b^\parall} \oplus \mathbf{\mathbf b^\perp}|^2= \frac{24}{25}(5-\sqrt5)\pi^2$ respectively.
For these dislocations, condition \eqref{eq:geocond} holds, meaning that dislocations nucleating at interfaces between misoriented or mismatched grains enjoy the property $\mathbf v^{\parall}=\mathbf v^{\perp}$. In this regime, from Eq.~\eqref{eq:vel_constraint}, we get an expression for the unique defect velocity
\begin{equation}\label{eq:velocity_equal}
     v_j 
    = \frac{8\pi^2}{5(|\mathbf b^\parall|^2+a^2|\mathbf b^\perp|^2)} \sum\limits_{n=1}^5 \frac{J_{n,j}^{\rm D}}{D_n}\Big|_{\mathbf{r}=\mathbf{0}} 
    = \frac{1}{2} \sum\limits_{n=1}^5 \frac{J_{n,j}^{\rm D}}{D_n}\Big|_{\mathbf{r}=\mathbf{0}},
\end{equation}
where we used $|\mathbf b^\parall|^2+a^2|\mathbf b^\perp|^2 = 16\,\pi^2 /5$ for the dislocations in question.
Evaluating the density current $\mathbf{J}_n^{\rm D}$ at $\mathbf{r}_0$ means evaluating 
$\partial_t \eta_n|_{\rv=\rv_0}\approx A\mathcal{G}^2\eta_n |_{\rv=\rv_0}$ 
from Eq.~\eqref{eq:dynamicseta} with $\zeta_k|_{\rv=\rv_0}=0$ due to amplitudes
$\eta_n$ vanishing at the core for $s_n \neq 0$ \cite{SKOGVOLL2022104932}.
By approximating the singular part of the phase $\theta_n$ with the isotropic vortex ansatz $s_n \arctan(y/x)$ \cite{Mazenko97,salvalaglioPRL2021,Skaugen2018b}, we get  
\begin{equation}\label{eq:zeroAmpVelocity}
\begin{split}
    \frac{J_{n,j}^{\rm D}}{D_n}\Big|_{\mathbf{r}=\mathbf{0}} &= 
    \frac{8A}{s_n} \epsilon_{jl} G^\parall_{n,l} G^\parall_{n,m} 
    \partial_m\theta_n \Big|_{\mathbf{r}=\mathbf{0}} 
    \\
    &= \frac{8A}{s_n} \epsilon_{jl} G^\parall_{n,l} G^\parall_{n,m}
    (G^\parall_{n,o} \partial_m u_o + G^\perp_{n,o}\partial_m w_o)
    \Big|_{\mathbf{r}=\mathbf{0}},    
\end{split}
\end{equation}
where we used Eq.~\eqref{eq:phases} to express the phase in terms of displacement fields.
Then, we can rewrite Eq.~\eqref{eq:velocity_equal} as
\begin{equation}\label{eq:vel_t}
\begin{split}
    v_j &= 4A \epsilon_{jl}\sum\limits_{n=1}^5 \frac{1}{s_n}  
    G^\parall_{n,l}G^\parall_{n,m}(G^\parall_{n,o}\partial_{m} u_o 
    +G^\perp_{n,o}\partial_{m} w_{o})\Big|_{\mathbf{r}=\mathbf{0}} 
    \\
    &= 4A \epsilon_{jl}\bigg[ 
    \left(\sum\limits_{n=1}^5 \frac{G^\parall_{n,m}}{s_n}\right)
    \sigma^{\parall}_{lm}  
    +\left(\sum\limits_{n=1}^5 \frac{G^\perp_{n,m}}{s_n}\right)
    \sigma^{\perp}_{lm} \bigg]_{\mathbf{r}=\mathbf{0}}. 
\end{split}
\end{equation}
By evaluating the sums
\begin{equation}
\begin{split}
    \sum\limits_{n=1}^5 \frac{G^\parall_{n,m}}{s_n} &= \frac{5}{4\pi} b^\parall_m,
    \\
    \sum\limits_{n=1}^5 \frac{G^\perp_{n,m}}{s_n} &= \frac{5a^2}{4\pi} b^\perp_m,
\end{split}
\end{equation}
we finally obtain a Peach-Koehler (PK) type equation
\begin{equation}\label{eq:vel}
    v_i = Mf_i^{\rm PK} =\frac{5A}{\pi} \epsilon_{ij}\left( \sigma^{\parall}_{jk}b_k^{\parall}+a^2\sigma^{\perp}_{jk}b_k^{\perp} \right),
\end{equation}
with $M=5A/\pi$, retaining a dependence on both phononic and phasonic deformation consistent with classical theories \cite{Lubensky86dislomotion,Agiasofitou_2010}. 

\begin{figure}
    \centering
    \includegraphics{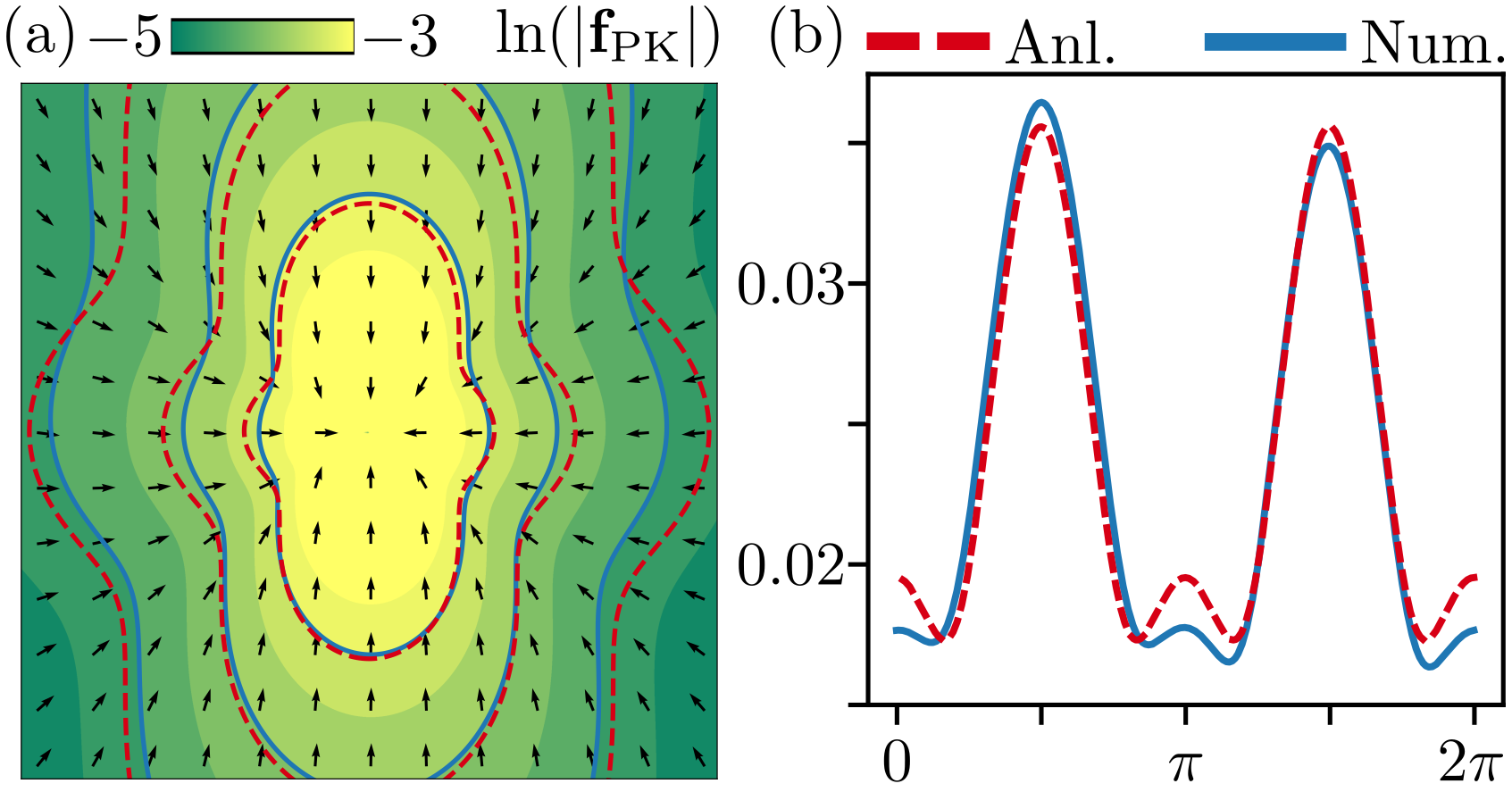}
    \caption{   
    \textit{Peach-Koehler force for selected dislocations in a decagonal QC}.
    (a) $\mathbf{f}_{\mathrm{PK}}$ from Eq.~\eqref{eq:vel}, generated by a dislocation at the center of the panel on a test dislocation with opposite Burgers vector. Direction (arrows) and magnitude (filled contour plot and solid isolines) are obtained with the stress field computed from amplitudes via Eq.~\eqref{eq:stressAPFC} (\textit{Num.}). Dashed isolines are obtained with Eq.~\eqref{eq:stressANL} exploiting analytic stress fields for dislocations reported in the Appendix \ref{app:analyticstress} (\textit{Anl.}). (b) Comparison between \textit{Num.} (solid) and \textit{Anl.} (dashed) $|\mathbf{f}_{\mathrm{PK}}|$ varying the polar angle at a distance $5 \Lambda$ from the center of panel (a).
    }
    \label{fig:figure4}
\end{figure}

An example of the PK force field evaluation is illustrated in Fig.~\ref{fig:figure4}. We compute $\mathbf{f}_{\mathrm{PK}}$ acting one a test dislocation with Burgers vector $\bv^{\parall} =  \bv^{\perp} = [0,\frac{8}{5}\pi \sin(\frac{4}{5}\pi)]$ and that is induced by stress field of an opposite-charged dislocation located at the origin. 
We report the result obtained by using the stress field computed from $\{\eta_n\}$ via Eq.~\eqref{eq:stressAPFC} (\textit{numerical}) and, for comparison, 
the prediction exploiting analytic stress field derived from the displacements in Ref.~\cite{PialiPRB1987} (\textit{analytic}); see full expressions in the Appendix \ref{app:analyticstress}.
Given a distribution of dislocations, we can then compute the PK force field based only on the symmetry of the QCs. Moreover, the compelling agreement between the \textit{numerical} and \textit{analytic} force fields further supports the consistency of the elasticity description for QCs achieved in the proposed theory with continuum mechanics while simultaneously capturing mesoscale aspects like defect formation and interaction with no additional tuning parameters. 

\section{Conclusion}
\label{sec:conclusions}

In summary, the proposed mesoscale field theory builds on the density-wave description of QCs. It focuses on the slowly-varying complex amplitudes of the characteristic Fourier modes of the microscopic density field. A free energy functional for these amplitudes is introduced. This newly proposed theory may be considered a mesoscale Landau theory for phase transition in QCs, including mechanics. Elasticity and dislocations, including phononic and phasonic deformations, follow from the symmetry of the microscopic quasicrystalline order and are shown to be consistent with classical continuum mechanics results. Using this approach, we shed light on the formation of semi-coherent interfaces between misoriented QCs.
Dislocation kinematics is also shown to follow from the proposed equation of motion for the amplitude. A model closing the gap between micro- and macroscopic description of QCs is thus established. We expect this theory to pave the way for general mesoscale investigations of systems with quasicrystalline order.

\section{Acknowledgements}

M. D. D. and M. S. acknowledge 
funding by the Deutsche Forschungsgemeinschaft (DFG, German Research
Foundation) Project No.~447241406, and the computing time made available to them on the high-performance computer at the NHR Center of TU Dresden.  K. R. E. acknowledges support from the National Science Foundation (NSF) under Grant No.~DMR-2006456.

\appendix

\section{Tiling}
\label{app:tiling}
To obtain the tiling shown in Fig.~\ref{fig:figure1} and \ref{fig:figure2}, we start from the density field reconstructed from the amplitudes according to Eq.~\eqref{eq:psir}. 
We numerically find all density maxima, and we discard those under an arbitrary threshold. We found that discarding maxima for which $\psi/\psi_{\rm max}<0.5$ matches best other tiling representations of quasicrystals (see, e.g., Ref.~\cite{Achim2014}). The coordinates of the density maxima remaining after thresholding are then the coordinates of the vertices we plot.

The length of the tile edge is once again arbitrary, as the QC presents a distribution of interatomic distances with multiple sharp peaks in non-commensurate positions. In our description, a tiling matching those in literature, with edges connecting most neighboring vertices and with no crossing edges, is obtained by choosing an edge length $l = 7.7 \pm 0.1$. The same value was used for the relaxed QC in Fig.~\ref{fig:figure1}  and the QC hosting defects in \ref{fig:figure2}, resulting in broken edges for the latter.

\section{Numerical method}
\label{app:numerics}

We numerically solve Eq.~\eqref{eq:dynamicseta} for each amplitude $\eta_n$.  
The first term on the right-hand side consists of an operator $\mathcal{O}$ linear in the amplitude, while the second term is a non-linear polynomial $\mathcal{N}$. 
We can thus rewrite it as
\begin{equation}\label{eq:fft1}
    \partial_t \eta_n = \mathcal{O}\eta_n + \mathcal{N}.
\end{equation}
This equation is solved by a Fourier pseudo-spectral method \cite{salvalaglio2022coarse}. In brief, we may consider the (discrete) Fourier transform of terms in Eq.~\eqref{eq:fft1} and rewrite the equation for the coefficient of the Fourier modes. This results in the equation
\begin{equation}
    \partial_t[\widehat{\eta}_n]_k = 
    {\mathcal{O}}_k[\widehat{\eta}_n]_k + [\widehat{\mathcal{N}}]_k,
\end{equation}
where $[\widehat{\eta}_n]_k$ is the Fourier transform of the amplitudes, ${\mathcal{O}}_k$ is a linear term consisting of a simple algebraic expression of the Fourier space coordinates $\mathbf{k}$ (for instance, $[\widehat{\nabla^2 \eta_n}]_k = -|\mathbf{k}_n|^2 [\widehat{\eta}_n]_k$), and $[\widehat{\mathcal{N}}]_k$ the Fourier transform of ${\mathcal{N}}$.
Knowing $\eta_n$ at time $t$, and thus its Fourier transform as well as $[\widehat{\mathcal{N}}]_k$, the amplitudes at the next timestep $t+\Delta t$ are given by the following approximation
\begin{equation}
    [\widehat{\eta}_n]_k (t+\Delta t) \approx 
    [\widehat{\eta}_n]_k (t) e^{{\mathcal{O}}_k\Delta t} + 
    \frac{[\widehat{\mathcal{N}}]_k(t)}{\mathcal{O}_k}
    \left(e^{{\mathcal{O}}_k\Delta t}-1\right). 
\end{equation}
We remark that the linear term is exact, while the approximation follows from evaluating the nonlinear part.
The solution in real space is then obtained by an inverse Fourier transform of $[\widehat{\eta}_n]_k$.
Our code is implemented in \texttt{python}, and it exploits the established Fast Fourier Transform algorithm FFTW, see also Ref.~\cite{Benoit-Marechal_2024}.
We use a timestep $\Delta t = 1$ and a uniform grid with ten mesh points per coarsening length $\Lambda$.

\section{Definitions of direct- and reciprocal-space vectors for decagonal quasicrystals}\label{app:geo}

Following seminal works for the decagonal quasicrystal \cite{Levine85elasticity}, we define five (four-dimensional) vectors $\mathbf{\widetilde{b}}_m$ belonging to the hyperlattice $\mathcal{L}$, and five vectors $\mathbf{\widetilde{G}}_n$ belonging to its reciprocal hyperlattice as
\begin{equation}\label{eq:geo_vectors}
\begin{split}
    \mathbf{\widetilde{b}}_m &= \mathbf{b}_m^\parall \oplus b\mathbf{b}_m^\perp, \qquad m=0,...,4 \\
    \mathbf{\widetilde{G}}_n &= \mathbf{G}_n^\parall \oplus a\mathbf{G}_n^\perp, \qquad n=0,...,4
\end{split}
\end{equation}
where $\mathbf{b}_m^\parall$ and $\mathbf{b}_m^\perp$ are vectors in parallel and perpendicular space, and $\mathbf{G}_n^\parall$ and $\mathbf{G}_n^\perp$ are vectors in their respective reciprocal spaces, as introduced in Sect.~\ref{sec:dislo}.  
The coefficients $a,\,b$ allow these vectors to have different norms. Owing to the symmetry of decagonal QCs, the vectors in the parallel space (direct and reciprocal) are
\begin{equation}\label{eq:geo_vec_par}
    \begin{split}
        \mathbf{b}_m^\parall &= r\left[-\sin(2\pi m/5),\cos(2\pi m/5)\right], \\
        \mathbf{G}_n^\parall &= g\left[\cos(2\pi n/5),\sin(2\pi n/5)\right], 
    \end{split}
\end{equation}
with $r,\,g$ their respective lengths. The vectors in the perpendicular space (direct and reciprocal) can be constructed by a simple re-indexing of those defined for the parallel space:
\begin{equation}\label{eq:geo_vec_perp}
        \mathbf{b}_m^\perp = \mathbf{b}_{3m}^\parall, \qquad
        \mathbf{G}_n^\perp = \mathbf{G}_{3n}^\parall. 
\end{equation}
We remark that, due to the periodicity of the sinusoids used to construct the vectors in Eq.~(\ref{eq:geo_vec_par}), a modulo five operation is implied in all the indices in Eqs.~\eqref{eq:geo_vec_par} and \eqref{eq:geo_vec_perp}, i.e., $\mathbf{b}_{m+5i}^\parall = \mathbf{b}_m^\parall \, \forall i \in \mathbb{Z}$, and the same for the other vectors.
In order to determine the coefficients $a,\,b,\,g,\,r$, we can consider a definition of phases following from Eq.~\eqref{eq:ointphase}. 
\begin{equation}
\begin{split}
    \widetilde\theta_{n-m} &= {\mathbf{\widetilde{G}}}_n \cdot {\mathbf{\widetilde{b}}}_m \\  &= gr \big( \sin(2\pi(n-m)/5) + ab \sin(6\pi(n-m)/5)\big).
\end{split}
\end{equation}
By imposing 
\begin{equation}
    \begin{cases}
         \widetilde \theta_1 =  \widetilde \theta_4 = gr (\sin(2\pi/5) - ab\sin(4\pi/5)) \equiv 0, \\
         \widetilde \theta_2 =- \widetilde \theta_3 = gr (\sin(4\pi/5) + ab\sin(2\pi/5)) \equiv 2\pi,
    \end{cases} 
\end{equation}
which is not the unique choice but the one most commonly considered in the literature, we get
\begin{equation}\label{eq:geo_abgr}
    \begin{cases}
        ab = \frac{1}{2}(1+\sqrt{5}), \\[1ex]
        gr = \frac{2\pi}{5} \sqrt{10 - 2 \sqrt{5}}. 
    \end{cases}
\end{equation}
We choose to take $g=1$, meaning that the reciprocal lattice vectors in parallel space have unit length, as usual in amplitude phase field crystal models \cite{salvalaglio2022coarse}. We also take $b=1$, meaning that the direct lattice vectors have the same length in both subspaces. Accordingly, $a$ and $r$ have well-defined values determined by Eq.~(\ref{eq:geo_abgr}) with $g=b=1$.

\section{Energy variation}
\label{app:energystress}

The elastic energy is given by Eq.~\eqref{eq:elastic_energy}.
The differential operator entering this equation can be written as  
\begin{equation}
\mathcal G_n = \nabla^2 + 2\I (\mathbf G_n\cdot \nabla) \\= \partial_j (\partial_j +2\I G_{n,j}).
\end{equation}
The variational of the amplitudes with respect to an arbitrary phase variation $\delta \theta_n$ is 
\begin{equation}
   \eta'_n = \eta_ne^{-\I\delta\theta_n} \approx \eta'_n = \eta_n(1-\I\delta\theta_n), 
\end{equation}
so that
\begin{equation}
    \delta \eta_n = -\I \eta_n  \delta\theta. 
\end{equation}
For infinitesimal phase variations, we then have the following relations between derivatives: 
\begin{equation}
\begin{split}
    \partial_j\delta \eta_n =& -\I (\partial_j\eta_n)  \delta\theta_n -\I \eta_n(\partial_j\delta\theta_n)\nabla^2\delta \eta_n 
    \\
    =& -\I (\nabla^2\eta_n)  \delta\theta_n -2\I (\partial_j\eta_n)  (\partial_j\delta\theta_n) \mathcal G_n \delta \eta_n  
    \\
    =& -\I (\nabla^2\eta_n)  \delta\theta_n -2\I (\partial_j\eta_n)  (\partial_j\delta\theta_n) \\
    &+2 (G_{n,j}\partial_j\eta_n)  \delta\theta_n +2 \eta_n(G_{n,j}\partial_j\delta\theta_n).    
    \end{split}    
\end{equation}
By rearranging terms, we get 
\begin{equation}
\begin{split}
\mathcal G_n \delta \eta_n 
&= -\I(\mathcal G_n\eta_n)(\delta\theta_n)-2\I(\mathcal Q_{n,j}\eta_n)(\partial_j\delta\theta_n),   
\end{split}
\end{equation}
where $\mathcal Q_{n,j} = \partial_j+\I G_{n,j}$. 

The variational of the elastic energy with respect to amplitude variations can then be written as
\begin{equation}
\begin{split}
\delta\mathcal E =& A\sum\limits_{n=0}^4 \int d\mathbf r \left[(\mathcal G_n\eta_n)(\mathcal G^*_n \delta\eta_n^*)+ (\mathcal G_n\delta\eta_n)(\mathcal G^*_n \eta_n^*)\right]\\
=& 2A\sum\limits_{n=0}^4 \int d\mathbf r \,\Re\left[(\mathcal G_n^*\eta_n^*)(\mathcal G_n \delta\eta_n)\right]. 
\label{eq:varene}
\end{split}
\end{equation}
By using the identities reported above,  
\begin{equation}
\begin{split}
    (\mathcal G^*_n\eta_n^*)(\mathcal G_n \delta \eta_n) 
=&  -\I|\mathcal G_n\eta_n|^2(\delta\theta_n)\\ &-2\I(\mathcal G^*_n\eta_n^*)(\mathcal Q_{n,j}\eta_n)(\partial_j\delta\theta_n).        
\end{split}
\end{equation}
The first term of the resulting expression is purely imaginary. Therefore, only the second term contributes to \eqref{eq:varene}, and
\begin{equation}
\Re\left[(\mathcal G^*_n\eta_n^*)(\mathcal G_n \delta \eta_n)\right] 
=  2\Im\left[(\mathcal G^*_n\eta_n^*)(\mathcal Q_{n,j}\eta_n)\right](\partial_j\delta\theta_n).
\end{equation}
By inserting this last expression into Eq.~\eqref{eq:varene}, we obtain the expression reported in the main text, Eq.~\eqref{eq:var_elast_phase}.

We note that the general expression for the stress obtained by coarse-graining the stress of the PFC density \cite{salvalaglio2022coarse} includes higher order terms w.r.t Eq.~\eqref{eq:stressAPFC} obtained via the energy variation above (see Sect.~\ref{sec:elasticity}). However, it reduces to the fields considered here by neglecting the highest (fourth) order only, with the two equations thus delivering very similar estimates.

\section{Strain and rotation fields from amplitudes}
\label{app:strain}

To obtain expressions for strain and rotation fields as a function of amplitudes, we first determine phonon and phason displacement from the phases of amplitudes. To simplify the notation, we redefine these phases, first introduced in Eq.~\eqref{eq:phases}, as:
\begin{equation}\label{eq:supp:newphase}
    \theta_n = \mathbf{k}_n\cdot\mathbf{u} + a \mathbf{q}_n\cdot\mathbf{w}, 
\end{equation}
so that $\mathbf{k}_n=\Gv^{\parall}_n$, $\mathbf{q}_n=\Gv^{\perp}_n$, and $a = (1+\sqrt{5})/2$ the geometrical factor determined in Appendix \ref{app:geo}.

The components of the displacement field $u_x,u_y,w_x,w_y$, are computed by inverting equation \eqref{eq:supp:newphase}, extending a procedure established for periodic crystal \cite{SalvalaglioNPJ2019}. We consider four amplitudes indexed by four (different) indexes $l,m,n,o$ and rewrite the algebraic problem in matrix form
\begin{equation}\label{eq:deform_system}
\begin{vmatrix}
    \theta_l \\
    \theta_m \\
    \theta_n \\
    \theta_o 
\end{vmatrix}=
\begin{vmatrix}
    k_l^x & k_l^y & a\,q_l^x & a\,q_l^y\\
    k_m^x & k_m^y & a\,q_m^x & a\,q_m^y\\
    k_n^x & k_n^y & a\,q_n^x & a\,q_n^y\\
    k_o^x & k_o^y & a\,q_o^x & a\,q_o^y
\end{vmatrix}\cdot
\begin{vmatrix}
    u^x \\
    u^y \\
    w^x \\
    w^y 
\end{vmatrix},
\end{equation}
By solving this system of equations, we obtain expressions for the components of the displacement fields:

\begin{widetext}
\begin{equation}
\label{eq:displfirst}
\begin{split}
    u_x = \frac{1}{\kappa_0}\bigg(
    &\theta_o (
        k_l^y q_m^y q_n^x - q_l^y k_m^y q_n^x - k_l^y q_m^x q_n^y + q_l^x k_m^y q_n^y + q_l^y q_m^x k_n^y - q_l^x q_m^y k_n^y) + \\ 
    & \theta_m (
        k_l^y q_n^y q_o^x + q_l^y q_n^x k_o^y -  q_l^x q_n^y k_o^y -  q_l^y k_n^y q_o^x -  k_l^y q_n^x q_o^y +  q_l^x k_n^y q_o^y) + \\
    & \theta_l (
       q_m^y k_n^y q_o^x + k_m^y q_n^x q_o^y + q_m^x q_n^y k_o^y - 
       k_m^y q_n^y q_o^x -  q_m^x k_n^y q_o^y - q_m^y q_n^x k_o^y ) + \\
    & \theta_n (
      -k_l^y q_m^y q_o^x + q_l^y k_m^y q_o^x + k_l^y q_m^x q_o^y 
      - q_l^x k_m^y q_o^y - q_l^y q_m^x k_o^y + q_l^x q_m^y k_o^y) 
    \bigg) , 
\end{split}
\end{equation}
\begin{equation}
\begin{split}
    u_y = \frac{-1}{\kappa_0}\bigg(
    &\theta_o(k_l^x q_m^y q_n^x - q_l^y k_m^x q_n^x - k_l^x q_m^x q_n^y + q_l^x k_m^x q_n^y + q_l^y q_m^x k_n^x - q_l^x q_m^y k_n^x) + \\
    &\theta_m (k_l^x q_n^y q_o^x +  q_l^y q_n^x k_o^x +  q_l^x k_n^x q_o^y -  q_l^y k_n^x q_o^x -  q_l^x q_n^y k_o^x -  k_l^x q_n^x q_o^y ) + \\
    & \theta_l (q_m^x q_n^y k_o^x + q_m^y k_n^x q_o^x + k_m^x q_n^x q_o^y -  q_m^x k_n^x q_o^y - k_m^x q_n^y q_o^x - q_m^y q_n^x k_o^x) + \\
    & \theta_n (-k_l^x q_m^y q_o^x + q_l^y k_m^x q_o^x + k_l^x q_m^x q_o^y -
      q_l^x k_m^x q_o^y - q_l^y q_m^x k_o^x +  q_l^x q_m^y k_o^x)\bigg),
\end{split}
\end{equation}
\begin{equation}
\begin{split}
    w_x = \frac{1}{a\kappa_0}\bigg(
    &\theta_o (k_l^y k_m^x q_n^y - k_l^x k_m^y q_n^y - k_l^y q_m^y k_n^x + q_l^y k_m^y k_n^x + k_l^x q_m^y k_n^y - q_l^y k_m^x k_n^y) + \\
    & \theta_m ( k_l^y k_n^x q_o^y + q_l^y k_n^y k_o^x + k_l^x q_n^y k_o^y 
      - k_l^x k_n^y q_o^y - k_l^y q_n^y k_o^x - q_l^y k_n^x k_o^y ) + \\
    & \theta_l ( k_m^x k_n^y q_o^y + k_m^y q_n^y k_o^x + q_m^y k_n^x k_o^y 
      - k_m^y k_n^x q_o^y - q_m^y k_n^y k_o^x - k_m^x q_n^y k_o^y ) + \\ 
    & \theta_n (-k_l^y k_m^x q_o^y + k_l^x k_m^y q_o^y + k_l^y q_m^y k_o^x - q_l^y     k_m^y k_o^x - k_l^x q_m^y k_o^y + q_l^y k_m^x k_o^y)\bigg), 
\end{split}
\end{equation}
\begin{equation}
\begin{split}
    w_y = \frac{1}{a\kappa_0}\bigg(
    & \theta_o (-k_l^y k_m^x q_n^x + k_l^x k_m^y q_n^x + k_l^y q_m^x k_n^x 
        - q_l^x k_m^y k_n^x - k_l^x q_m^x k_n^y + q_l^x k_m^x k_n^y) + \\
    & \theta_m ( k_l^x k_n^y q_o^x + k_l^y q_n^x k_o^x + q_l^x k_n^x k_o^y 
        - k_l^y k_n^x q_o^x - q_l^x k_n^y k_o^x - k_l^x q_n^x k_o^y) + \\ 
    & \theta_l ( k_m^y k_n^x q_o^x + q_m^x k_n^y k_o^x + k_m^x q_n^x k_o^y 
        - k_m^x k_n^y q_o^x - k_m^y q_n^x k_o^x - q_m^x k_n^x k_o^y ) + \\ 
    & \theta_n (k_l^y k_m^x q_o^x - k_l^x k_m^y q_o^x - k_l^y q_m^x k_o^x 
        + q_l^x k_m^y k_o^x + k_l^x q_m^x k_o^y - q_l^x k_m^x k_o^y)\bigg), 
\end{split}
\label{eq:displlast}
\end{equation}
\end{widetext}
with
\begin{equation}
    \kappa_0 = \mathrm{det}
        \begin{vmatrix}
            k_l^x & k_l^y & q_l^x & q_l^y\\
            k_m^x & k_m^y & q_m^x & q_m^y\\
            k_n^x & k_n^y & q_n^x & q_n^y\\
            k_o^x & k_o^y & q_o^x & q_o^y
        \end{vmatrix}.
\end{equation}

We remark that for the decagonal QC described via five amplitudes, one can construct five solutions to Eq.~(\ref{eq:deform_system}) as only four amplitudes are needed; however, these solutions are all equivalent.   

Expressions for the strain fields $\varepsilon_{ij}^{\parall}= \frac{1}{2}(\partial_iu_j + \partial_ju_i)$ and $\varepsilon_{ij}^{\perp}=\partial_iw_j$ follow from the expression of the displacements. 
The space-dependence of the displacements obtained by solving Eq.~(\ref{eq:deform_system}) is fully contained in the phases $\theta_j$. 
Therefore, spatial derivatives of displacements can be computed by using expressions analogous to \eqref{eq:displfirst}--\eqref{eq:displlast} featuring $\partial_i \theta_n$ terms instead of $\theta_n$. Derivatives of the phases can be generally computed as
\cite{SalvalaglioNPJ2019}
\begin{equation}
    \frac{\partial \theta_j}{\partial x_i} = \frac{1}{|\eta_j|^2} \bigg(
    \frac{\partial \mathrm{Im}(\eta_j)}{\partial x_i} \mathrm{Re}(\eta_j) -
    \frac{\partial \mathrm{Re}(\eta_j)}{\partial x_i} \mathrm{Im}(\eta_j)
    \bigg),
\end{equation}
from which strain tensor components can be expressed in terms of the amplitudes and reciprocal-lattice vectors only. For instance,
\begin{widetext}
\begin{equation}
\begin{split}
    \varepsilon^{||}_{xx}=\partial_x u_x = \frac{1}{\kappa_0}\bigg(
    &\partial_x\theta_o (
        k_l^y q_m^y q_n^x - q_l^y k_m^y q_n^x - k_l^y q_m^x q_n^y + q_l^x k_m^y q_n^y + q_l^y q_m^x k_n^y - q_l^x q_m^y k_n^y) + \\ 
    &\partial_x \theta_m (
        k_l^y q_n^y q_o^x + q_l^y q_n^x k_o^y -  q_l^x q_n^y k_o^y -  q_l^y k_n^y q_o^x -  k_l^y q_n^x q_o^y +  q_l^x k_n^y q_o^y) + \\
    &\partial_x \theta_l (
       q_m^y k_n^y q_o^x + k_m^y q_n^x q_o^y + q_m^x q_n^y k_o^y - 
       k_m^y q_n^y q_o^x -  q_m^x k_n^y q_o^y - q_m^y q_n^x k_o^y ) + \\
    &\partial_x \theta_n (
      -k_l^y q_m^y q_o^x + q_l^y k_m^y q_o^x + k_l^y q_m^x q_o^y 
      - q_l^x k_m^y q_o^y - q_l^y q_m^x k_o^y + q_l^x q_m^y k_o^y) 
    \bigg),  
\end{split}
\end{equation}
\end{widetext}
while other components of the strain tensor can be obtained by proceeding analogously with other expressions and/or derivatives.
Note that the resulting strain components are continuously defined everywhere except at the core of dislocation, where, however, the effective elastic constants vanish (see also additional discussions for equations with a similar form in Ref.~\cite{SalvalaglioNPJ2019,salvalaglio2022coarse}).

Similarly, the rotation fields can be derived from amplitudes. 
Using the expressions for the displacement derived above, one can compute them according to the following definitions
\begin{equation}
    \omega^\parall = \curl \mathbf{u}, \qquad 
    \omega^\perp = \curl \mathbf{w}, 
\end{equation}
representing rotations in $\Omega^{\parallel}$ and $\Omega^{\perp}$ (planes), respectively. 
For instance, to plot the rotation field $\omega \equiv \omega^{\parall}$ illustrated in Fig.~\ref{fig:figure2}, we selected the amplitude indices ${\{l,m,n,o\}} = {\{1,2,3,4\}}$. In this case, explicitly reporting the values of products of reciprocal-space vector components for convenience, the rotation fields result:   
\begin{equation}
\begin{split}
    \omega^\parall \sim & \, 
    0.38~\partial_x(\theta_1 + 0.618 \theta_2 
          - 0.618 \theta_3 - \theta_4 ) \\
          &+
    0.276~\partial_y(\theta_1 + 2.618 \theta_2
          + 2.618 \theta_3 + \theta_4), \\
    \omega^\perp \sim & \,
    0.145~\partial_x(\theta_1 + 1.618 \theta_2 - 1.618 \theta_3 - \theta_4) \\
        & + 
    0.447~\partial_y(\theta_1 + 0.382 \theta_2 + 0.382 \theta_3 + \theta_4 ).
\end{split}
\end{equation}

\begin{figure*}
    \centering
    \includegraphics{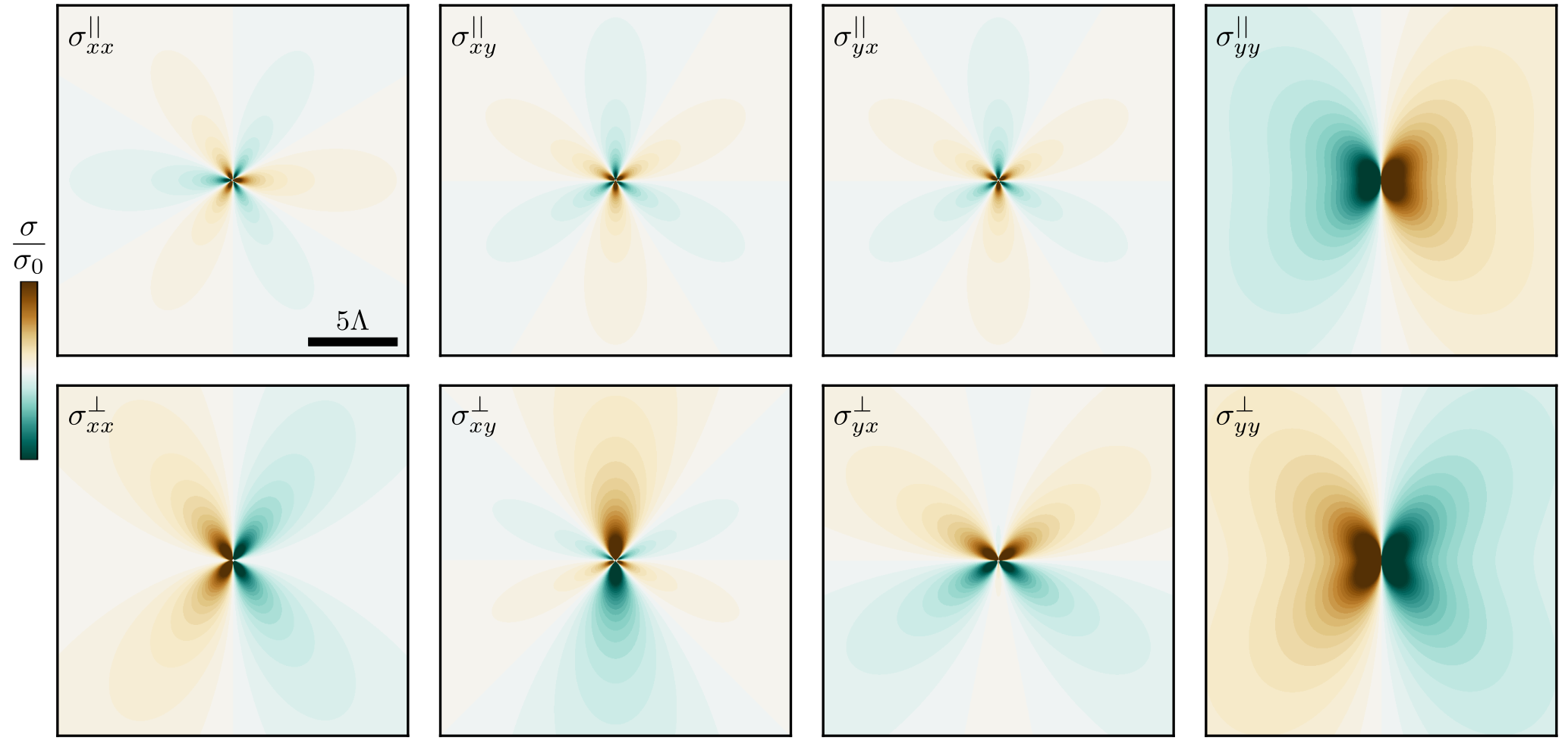}
    \caption{Components of the analytic stress field, Eqs.~\eqref{eq:supp:sigParANL} and \eqref{eq:supp:sigPerpANL}, for a dislocation with the Burgers vector observed along the small-angle grain boundary in Fig.~\ref{fig:figure3}. 
    The Burgers vectors are: 
    $\bv^{\parall} = \left[0,\frac{2\pi}{5} \sqrt{2 (\sqrt{5}+5)}\right],\,\bv^{\perp} = \left[0,-\frac{4\pi}{5} \sqrt{5-2 \sqrt{5}} \right]$. }
    \label{fig:supp::stressANL}
\end{figure*}

\section{Analytic stress fields for a dislocation in an elastically isotropic quasicrystal}
\label{app:analyticstress}

Analytic expressions for the displacement fields $\mathbf u$ and $\mathbf w$ of a dislocation in an elastically isotropic 2D QC (or in the plane perpendicular to a straight dislocation line in 3D) have been reported in Ref.~\cite{PialiPRB1987}. 
By computing the strains $\varepsilon^{\parall}_{ij}$ and and $\varepsilon^{\perp}_{ij}$ from such expressions 
and via the constitutive relations \eqref{eq:stressANL} we obtain the following \textit{analytic} expression of the stress field 
\begin{widetext}
\begin{equation}\label{eq:supp:sigParANL}
    \begin{split}
        \sigma_{xx}^\parall &=
        (2\lambda+3R)\bigg(\frac{b^\parall_y x - b^\parall_x y}{6 \pi  \left(x^2+y^2\right)}\bigg) +
        2\mu 
        \bigg( \frac{b^\parall_y x\left(x^2-3 y^2\right)-b^\parall_x y\left(5 x^2+y^2\right)}{6 \pi  \left(x^2+y^2\right)^2} \bigg),   
        \\
        \sigma_{xy}^\parall = \sigma_{yx}^\parall &= 
        2\lambda 
        \bigg( \frac{\left(x^2-y^2\right) (b^\parall_x x+b^\parall_y y)}{3 \pi  \left(x^2+y^2\right)^2} \bigg) 
        +R
        \bigg( \frac{b^\perp_x x + b^\perp_y y}{2 \pi  \left(x^2+y^2\right)} \bigg), 
        \\
        \sigma_{yy}^\parall &= 
        (2\lambda-3R)\bigg(\frac{b^\parall_y x - b^\parall_x y}{6 \pi  \left(x^2+y^2\right)}\bigg)   +
        2\mu
        \bigg( \frac{b^\parall_x y\left(3 x^2 -y^2\right)+b^\parall_y x\left(x^2+5 y^2\right)}{6 \pi \left(x^2+y^2\right)^2} \bigg), 
        \\
    \end{split}
\end{equation}
and
\begin{equation}
\begin{split}\label{eq:supp:sigPerpANL}
    \sigma_{xx}^\perp &= 
    -R
    \bigg( \frac{4xy(b^\parall_x x+b^\perp_y y)}{3 \pi  \left(x^2+y^2\right)^2} \bigg) 
    - K_1 
    \bigg( \frac{b^\perp_x y}{2 \pi  \left(x^2+y^2\right)} \bigg)
    + K_2 
    \bigg( \frac{b^\perp_y x}{2 \pi  \left(x^2+y^2\right)} \bigg),
    \\
    \sigma_{xy}^\perp &=   
    -2R 
    \bigg( \frac{\left(x^2-y^2\right) (b^\parall_x x+b^\parall_y y)}{3 \pi  \left(x^2+y^2\right)^2} \bigg)
    - K_1 
    \bigg( \frac{b^\perp_y y}{2 \pi  \left(x^2+y^2\right)} \bigg)
    + K_2
    \bigg( \frac{b^\perp_y y}{2 \pi  \left(x^2+y^2\right)} \bigg),
    \\
    \sigma_{yx}^\perp &=   
    2R 
    \bigg( \frac{\left(x^2-y^2\right) (b^\parall_x x+b^\parall_y y)}{3 \pi  \left(x^2+y^2\right)^2} \bigg) 
    + K_1
    \bigg( \frac{b^\perp_x x}{2 \pi  \left(x^2+y^2\right)} \bigg)
    - K_2
    \bigg( \frac{b^\perp_x x}{2 \pi  \left(x^2+y^2\right)} \bigg),
    \\
    \sigma_{yy}^\perp &=   
    -R  \bigg( \frac{4xy(b^\parall_x x+b^\perp_y y)}{3 \pi  \left(x^2+y^2\right)^2} \bigg) 
    + K_1 
    \bigg( \frac{b^\perp_y x}{2 \pi  \left(x^2+y^2\right)} \bigg)
    - K_2
    \bigg( \frac{b^\perp_x y}{2 \pi  \left(x^2+y^2\right)} \bigg).
\end{split}
\end{equation}
\end{widetext}

To compare with numerical simulations, we consider the elastic constants \eqref{eq:elconstants} in terms of $\lambda$, $\mu$, $R$, $K_{1,2}$ as obtained in Sect.~\ref{sec:elasticity}. In Fig.~\ref{fig:supp::stressANL}, we show the \textit{analytic} stress components for a dislocation of the same type as those observed along the small angle grain boundary in Fig.~\ref{fig:figure3} in the main text. It is found to match well the corresponding numerical simulation, similar to the prototypical case of the elastic field generated by a dislocation dipole (Fig.~\ref{fig:stressComp}).

\end{document}